# Numerical investigation of unsteady flow in a reversible pump-turbine

Chirag Trivedi

Associate Professor, Waterpower Laboratory, Department of Energy and Process Engineering, NTNU—Norwegian University of Science and Technology, Alfred Getz' vei. 4, 7491 Trondheim, Norway.
Email: chirag.trivedi@ntnu.no

Friday, 23 January 2026

ABSTRACT
Pump-turbine type hydraulic turbine is widely used to mitigate the intermittent energy demand and store a large-scale energy. Flow conditions in turbine mode and pump mode operations is substantially different. A computational model of the pump-turbine was created, and the model included hexahedral mesh of 58.19 million nodes. Total verification and validation error was 7.7%. Three operating conditions in turbine mode and four in pump mode were simulated. Flow characteristics, such as blade loading, time-dependent pressure fluctuations, frequency spectra, radial and tangential velocity were investigated. The frequency spectra revealed amplitude of frequencies up to tenth harmonics of the blade passing frequency in pump mode. The higher harmonic frequencies can potentially reach the high mode eigen frequencies and increase the risk of resonance. Flow field analysis in the draft tube indicted the strong presence of Dean vortices causing highly asymmetric flow at the runner inlet in pump mode operation.

# Introduction

Annual $CO_2$ emission has reached 42 billion tonnes and increases continuously. The trajectory has been largely parabolic since the year 1900, the annual $CO_2$ emission was around one billion tonnes. China, North America and Europe account for 53%, 18% and 17% of global emissions in recent years, respectively. The impact of $CO_2$ and greenhouse gases on climate is clearly visible. To minimize the impact, the concept of energy mix is being implemented worldwide, and the share of renewable energy has been steadily increasing from 18% in the year 2000 to 30% in 2023. The wind and solar energy integration is reached to 1000 GW in recent years. On the other hand, to provide stable energy to the consumers, flexibility and grid balance are essential. The gas turbine and hydropower plants play pivotal role to mitigate intermittency, from the wind and solar energy, and balance the power grid. The major limitation for the gas turbine power plant is the emission of carbon dioxide, nitrogen oxide and other organic compounds. The emission is high during the grid balance where the turbines operate at off design loads and results in abnormal combustion. Hydropower is recognized as a valuable alternative energy source due to its substantially lower emissions. The global energy production from hydropower is around 12 000 TWh. The hydropower provides high flexibility, and it can operate in tandem with solar and wind power generators.

Reversible pump-turbine is an essential component of an energy storage type hydropower plant. The pump-turbine generates electricity in turbine mode and store energy in pump mode operations depending on requirement. Therefore, the pump-turbine plays critical role in the energy system to mitigate flexibility. The rotational direction is opposite in pump mode and the water flows in reverse direction, i.e., from the lower reservoir to the upper reservoir. The pump-turbine is integrated with other valuable components, such as, spiral casing to guide the water, stay vanes to distribute the water circumferentially, guide vanes to control the water flow through the turbine, runner to extract energy from the water, and draft turbine to recover the pressure energy. Modern pump-turbines have achieved the efficiency up to 94% both turbine and pump mode operations at design load [1]. However, at off design loads the efficiency is low, a round-trip efficiency drops below 50%. There are several challenges to operate pump-turbine at off-design loads [2], however, the key challenge corresponds to the unsteady vortex breakdown and the resulting dynamic instability [3]. Therefore, the operating range is restricted and, as a result, the limited scope for mitigating the energy flexibility.

At design load, flow separation is minimal in the blade channel as the guide vane outlet angle, and the blade inlet angle are well aligned. At high load and part load conditions, the flow angles are skewed and induce asymmetric flow to the blade leading edge. This results in flow separation around the leading edge, and the separation grows as flow travels further downstream in the blade channel. Furthermore, angular rotation of the turbine runner, induce additional effect on the flow causing inception of vortical flow and the local swirls.

Reversible pump-turbine is stable and works well around the design load [4–7]. There is moderate impact on the turbine due to interaction between the rotating component (runner/impeller) and the stationary component (guide vanes and stay vanes). This rotor-stator interaction is investigated by the researchers [8,9] and, generally, the turbines withstand this type of repetitive dynamic loads at design condition. At off-design load, the turbine experiences challenges, high pressure amplitudes from vortex breakdown in the blade channel, draft tube and vaneless space [10–12].

The blade channels are subjected to different categories of vortical flow, depending on the operating conditions of the reversible pump-turbine. So far, the emphasize has been placed on the s-curve characteristics and the rotating stall. However, inception of vortical flow during other operating conditions is major challenge due to loss of efficiency, and it is not clearly understood, specifically off-design loads. Furthermore, there is no or very limited study on the pump mode operation of the turbine. The flow field in the blade channel is significantly different from that of the turbine mode. This study aims to investigate the unsteady flow in the vaneless space and runner blade channel of the reversible pump-turbine at wide operating range. The study encapsulates both modes of operation, i.e.,

C Trivedi

pump and turbine. Furthermore, we investigated, blade loading and other secondary flow characteristics during the inception of unsteady vortical flow, Dean vortices, in the pump mode operation. According to the author's the best of knowledge, no research on pump mode is available in the literature; although, the pump mode operation is important to store energy. The present study provides valuable information on the flow characteristics in pump mode operation of a pump-turbine.

# Computational model and range of parameters

## Reversible pump-turbine

The present study is conducted on a reversible pump-turbine test facility in the Waterpower laboratory, NTNU, Norway. The turbine is a reduced scale model of a prototype in Norway. The test facility is used for the research and model tests of the model Francis turbine and the reversible pump-turbine. Detailed information about the test facility is presented in our previous studies [13–15]. The reversible pump-turbine includes 6 blades, 28 guide vanes, 14 stay vanes, spiral casing and elbow type draft tube. The inlet and outlet diameter of the model runner are 0.631 m and 0.349 m, respectively. The maximum efficiency at the design load is 90.5%. Other reference parameters are presented in **Table 1**. An isometric view of the turbine is shown in **Figure 1**. The computational model of the turbine with all components was considered for the numerical modelling. **Figure 2** shows the final mesh considered for the present study. Hexahedral mesh was created in all components. ANSYS® ICEM CFD™ was used for creating the hexahedral mesh in the spiral casing and draft tube domains. ANSYS® TGrid™ was used for creating the hexahedral mesh in the guide vane and the runner domains. The final mesh contained 58.19 million nodes after the proper verification and validation.

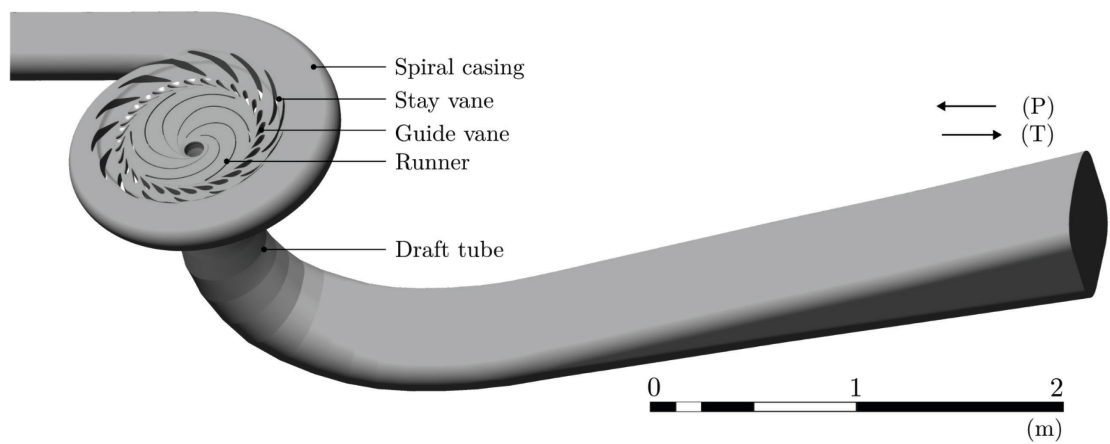

**Figure 1** Computational model of the turbine considered for the present study. The flow directions (P) and (T) indicate the pump and turbine mode operations, respectively.

Table 1 Overall design data of the investigated pump-turbine.

| Parameter | Value |
| --- | --- |
| Runner inlet diameter ($D_1$) | 0.631 m |
| Runner outlet diameter ($D_2$) | 0.349 m |
| Runner inlet height ($B_1$) | 0.059 m |
| Runner specific speed ($N_q$) | 27.1 (rpm m$^{3/4}$ s$^{-1/2}$) |
| Angular speed ($n*$) | 10.8 rev s$^{-1}$ |
| Flow rate ($Q*$) | 0.275 m$^3$ s$^{-1}$ |
| Head ($H^*$) | 29.3 m |
| Inlet blade angle ($\beta_1$) | 12° |
| Outlet blade angle ($\beta_2$) | 12.8° |
| Guide vane angle ($\alpha*$) | 10° |
| Speed factor ($N_{ED}$) | 0.223 |
| Discharge factor ($Q_{ED}$) | 0.133 |

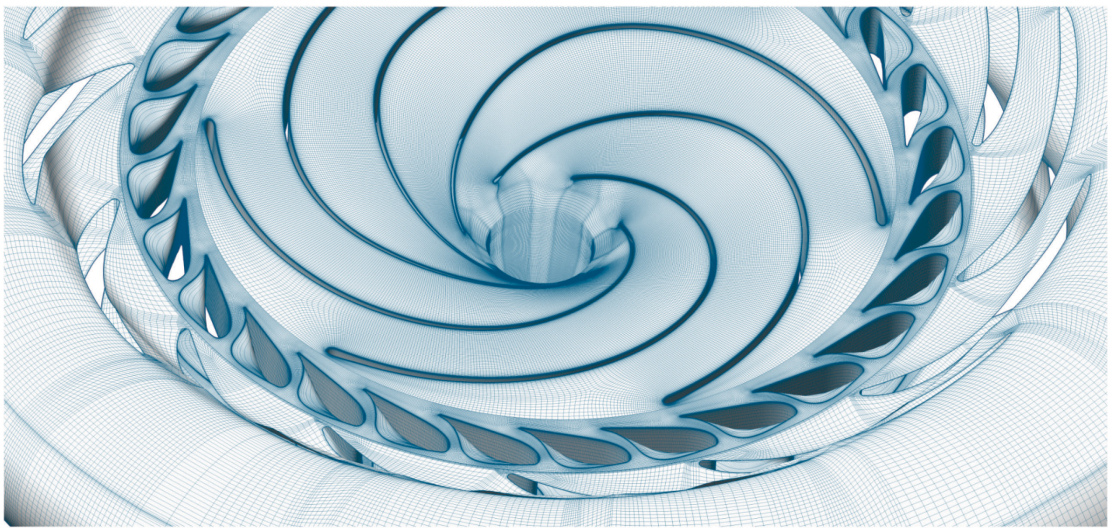

**Figure 2** Hexahedral mesh in the computational domain of the pump-turbine.

The simulations were carried out using ANSYS® CFX™. A high-performance computing cluster available at NTNU was used for the simulations. The cluster took around one million CPU hours to complete the simulations. Steady state simulations were conducted at all load conditions initially. Reynolds-averaged Navier-Stokes (RANS) based shear stress transport (SST) turbulence model was used [16]. The steady state results of SST were used to initialize the unsteady simulations at the corresponding loads. RANS modelling approaches do not provide essential information of the unsteady vortex due to averaging of the fluctuating component and minimum resolution of the turbulent length scales [17–19]. For resolving maximum possible turbulent length scales, a large eddy simulation (LES) is a natural choice, specifically turbulent flow in the blade channels [20]. However, LES requires fine mesh (order of trillion node points) considering the Reynolds numbers in a pump-turbine. Pacot et al. [21] estimated the mesh requirement in a pump-turbine is 900 x 10⁹ nodes for a LES. This is highly expensive and time consuming. Alternatively, a scale-adaptive simulations (SAS-SST) is an optimal turbulence modelling approach, taking the benefit of local fine mesh and dynamically adjusts the resolved length scales, Artificial Forcing function and adjusting the von Karman length scale. A detailed information about the mathematical model of SAS-SST is available in the literature [22,23]. SAS tends to an explicit transfer of kinetic energy, where the flow instabilities are not at the level of LES requirement. SAS uses forcing terms in the momentum equation to convert into resolved kinetic energy from the modelled kinetic energy [23]. SAS-SST model has demonstrated the success of modelling the complex flow conditions in hydraulic turbines including the reversible pump-turbine [11,24,25].

$$Q_{SAS} = \max\left[\rho\zeta_2 S^2 \left(\frac{L}{L_{vk}}\right)^2 - C_{SAS}\left(\frac{2\rho k}{\sigma_\phi}\right)\max\left(\frac{1}{k^2}\frac{\partial k}{\partial x_j}\frac{\partial k}{\partial x_j}, \frac{1}{\omega^2}\frac{\partial \omega}{\partial x_j}\frac{\partial \omega}{\partial x_j}\right), 0\right]; \quad (1)$$

where $\zeta_2$ = 1.47, $\sigma_\varphi$ = 2/3 and $C_{SAS}$ = 2.

$$L = c_\mu^{-\frac{1}{4}} \frac{\sqrt{k}}{\omega} \quad (2)$$

$$L_{vk} = \kappa \left|\frac{U'}{U''}\right| \quad (3)$$

$$U' = \sqrt{2 S_{ij} S_{ij}} \quad (4)$$

$$U'' = \sqrt{\frac{\partial^2 U_i}{\partial x_k^2}\frac{\partial^2 U_i}{\partial x_j^2}} \quad (5)$$

A high wave number damping is considered to minimize the accumulation of energy at the smallest scale. This is achieved by imposing the lower limit of the von Karman length scale as,

$$L_{vk} = \max\left(\kappa \left|\frac{U'}{U''}\right|; C_s \sqrt{\frac{\kappa \zeta_2}{\frac{\beta}{c_\mu} - \alpha}} \Delta\right); \text{ where } \Delta = \sqrt[3]{V}, \quad (6)$$

where, $C_s$ = 0.11, $\kappa$ = 0.41, $c_\mu$ = 0.09 and $\alpha$ = 0.44, $\beta$ = 0.0828 are coefficients for SST model.

Given the available computational resources and suitability for hydraulic turbines, the SAS–SST turbulence model was selected as the optimal approach for this study. All unsteady simulations were therefore performed using the SAS–SST turbulence model. The selected time-step size for the present study was one degree of runner rotation. The researchers [9,25–30] used 0.5 – 2 degree time-step size depending on requirement for the residual convergence. The simulations were performed for a total of 10 revolutions of the runner at each operating points to ensure converged flow field. The General Grid Interface between the rotating and stationary domain was modelled as transient rotor-stator interface.

This enables the flow variables to exchange information at every time-step and updates the angular position of the rotating domain accordingly. This interface requires extra simulation time as compared to the frozen rotor interface type. The inlet and outlet boundary conditions were mass flow inlet and pressure outlet types for both pump and turbine mode operations. Available experimental data [31] were used to prescribe the boundary condition. The inlet and outlet boundaries were swapped for pump mode operation. The inlet boundary corresponds to the inlet of the draft tube, and the outlet boundary corresponds to the outlet of the spiral casing. Additionally, the rotational direction of the runner was reversed.

## Range of parameters

The simulations were conducted at three operating points in turbine mode and four in pump mode. **Table 2** shows the overall parameter at different load conditions of the turbine. The parameters are obtained from the measurements. The reference design load corresponds to $\alpha = 1$ for both turbine and pump modes. High load and part load correspond to the $\alpha = 1.18$ and $\alpha = 0.64$, respectively. For pump mode, full load, intermediate load and part load correspond to $\alpha = 1.3$, $\alpha = 1.18$ and $\alpha = 0.64$, respectively. The maximum hydraulic efficiency was obtained at $\alpha = 1.3$.

**Table 2** Range of parameters investigated in turbine and pump mode operation.

| $\alpha$ | $n$ | $Q$ | $p_2$ | $H$ | $\eta_h$ |
|---|---|---|---|---|---|
| Turbine mode | | | | | |
| 0.64 | 679 | 0.184 | 53 439 | 29.9 | 0.866 |
| 1.00 | 679 | 0.281 | 53 110 | 29.9 | 0.909 |
| 1.18 | 679 | 0.316 | 50 188 | 29.9 | 0.906 |
| Pump mode | | | | | |
| 0.64 | 510 | 0.140 | 195 240 | 11.9 | 0.688 |
| 1.00 | 510 | 0.161 | 201 230 | 11.9 | 0.755 |
| 1.18 | 510 | 0.165 | 196 090 | 11.9 | 0.768 |
| 1.30 | 510 | 0.174 | 188 520 | 11.9 | 0.807 |

## Verification and validation

Verification and validation of the computational model was carried out before proceeding the final simulations. The verification was carried out using methods presented in the literature for hydraulic turbines [32]. The grid convergence method (GCI) is used for the estimation and reporting of uncertainty due to discretization in CFD applications [33]. This method is primarily based on Richardson extrapolation method [34], in addition, it takes advantage of extrapolated mesh and the variable values. **Table 3** shows an overview of the discretization error. Torque represents the global performance of a turbine therefore it was chosen for the verification. Numerical torque values for different mesh nodes are shown. The apparent order ($p$) is 1.668, which is near to the actual discretization scheme, second order upwind scheme, selected for the simulations.

$$\text{GCI}_{\text{fine}}^{21} \frac{1.25 e_a^{21}}{r_{21}^p - 1} \tag{7}$$

**Table 3** Overview of the numerical mesh and discretization error determined using GCI method.

| Variable | Value |
|---|---|
| $N_1$; $N_2$; $N_3$ | 40 566 564; 18 059 692; 8 016 432 |
| $r_{21}$; $r_{32}$ | 1.311; 1.310 |
| $\varphi_1$ | 1119.6 Nm |
| $\varphi_2$ | 1137.6 Nm |
| $\varphi_3$ | 1165.9 Nm |
| $p$ | 1.668 |
| $\varphi^{21}_{\text{ext}}$ | 1087.9 Nm |
| $e^{21}_a$ | 1.6% |
| $e^{21}_{\text{ext}}$ | 2.9% |
| $\text{GCI}^{21}_{\text{fine}}$ | 3.5% |

The fine contains 40.56 million nodes. The numerical uncertainty in the fine-grid solution for the torque value is 3.54%. More information on the GCI analysis and the verification is presented here [35]. The mesh was further improved, specifically around the blades in the runner. The final mesh selected for the simulations was 58.19 million nodes in the pump-turbine.

The validation of the numerical results was carried out with experimental data. **Figure 3** shows the iso-efficiency diagram of the model pump-turbine in turbine mode operation. The maximum measured efficiency of the turbine is $91.2^{\pm 0.2\%}$, $Q^*_{ED} = 1$ and $n^*_{ED} = 1$. **Figure 4** shows the performance characteristics in pump mode operation. The toque was considered for the numerical validation. **Table 4** shows comparison of numerical and experimental values for turbine and pump modes. Torque value represents the global order of accuracy and the important performance parameter for the blades. It indicates the performance of both stationary and rotating domains, spiral casing, guide vanes, runner and draft tube. The validation error is computed using **Equation 8**. The maximum validation error is obtained at $\alpha = 0.64$. in turbine mode. The minimum validation error is obtained at the design load ($\alpha = 1$). The flow field is stable, and the numerical model performs well at the design load. For the pump mode operation, the numerical model performs well across the selected loads, where the error is 9 – 10%. Hydraulic performance of the pump-turbine is nearly stable at these operating conditions.

C Trivedi

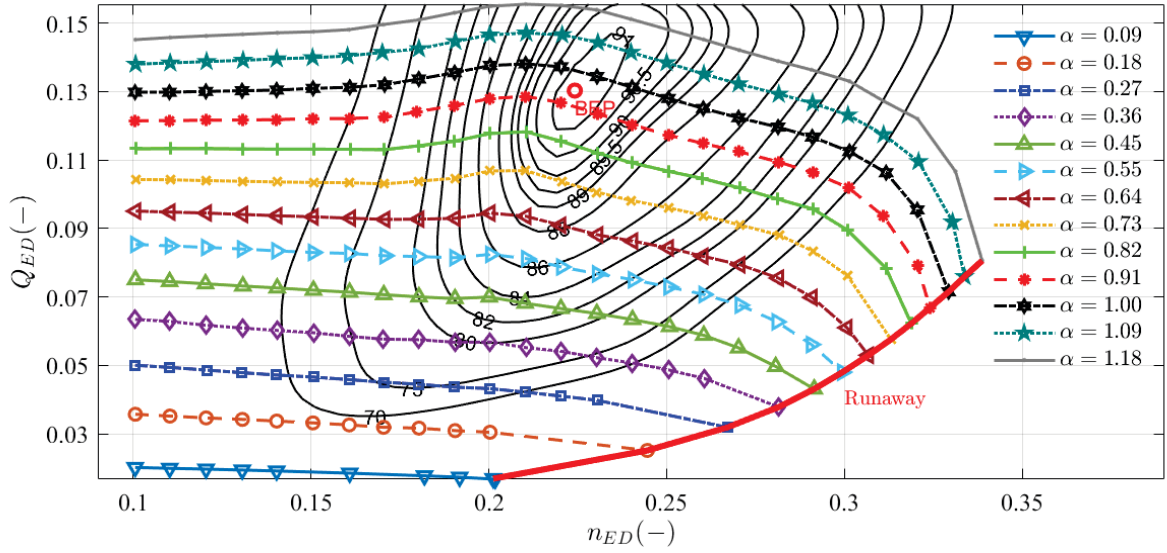

**Figure 3** Iso-efficiency diagram of the model pump-turbine considered for the present study [31].

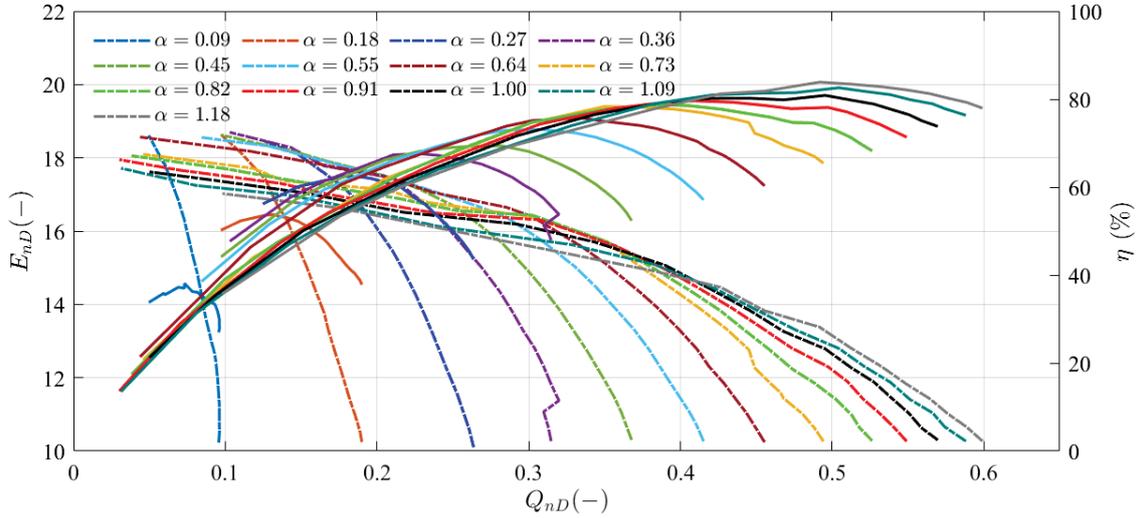

**Figure 4** Performance of the model pump-turbine in pump mode [31]. Dash lines represent $Q_{nD}$ and $E_{nD}$; continuous lines represent the efficiency on secondary y-axis.

For accurate verification and validation, it is important to consider the total error including the measurement uncertainties, discretization error and the validation error. **Equation 9** shows the computation of the total numerical error in torque value. The discretization error is 3.5%; the validation error is 5.6% and the experimental measurement error is 3.9%. The experimental measurement error includes, (1) uncertainties from the calibration of all instruments and sensors, (2) systematic and random error, (3) repeatability error. Detailed quantification of the experimental measurement error is presented in the literature [31]. Thus, the total error in the torque is 7.7% at the design load under turbine mode operation. The computational model of the pump-turbine

C Trivedi

demonstrated good performance in both turbine and pump mode operations. There was no unexpected and significant deviation in the global parameters.

Table 4 Validation of the numerical toque values at the selected operation points.

| α | $T_{exp}$ | $T_{num}$ | $\hat{e}_v$ (%) |
|---|---|---|---|
| Turbine mode | | | |
| 0.64 | 651.3 | 712.2 | 9.3 |
| 1.00 | 1049.1 | 1108.4 | 5.6 |
| 1.18 | 1198.2 | 1280.3 | 6.9 |
| Pump mode | | | |
| 0.64 | 453.3 | 409.3 | 9.7 |
| 1.00 | 471.7 | 429.0 | 9.1 |
| 1.18 | 474.8 | 427.3 | 10 |
| 1.30 | 479.3 | 435.9 | 9.1 |

$$\hat{e}_v = \frac{|T_{\text{num}} - T_{\text{exp}}|}{T_{\text{exp}}} \times 100 \quad (8)$$

$$\hat{e}_t = \sqrt{\hat{e}_{GCI}^2 + \hat{e}_v^2 + \hat{e}_{exp}^2} \quad (9)$$

# Results and discussions

Steady state and unsteady simulations were conducted, and the unsteady simulations were initialized with the results of the steady state simulations. The simulations correspond to three load values in turbine mode and four load values in pump mode. The simulations results are divided into two sections reflecting the turbine mode and pump mode operations of the reversible pump-turbine.

## Turbine mode

### Interaction of guide vane and blade

The turbine includes both stationary—spiral casing, stay vane, guide vane and draft tube—and rotating (runner, blade, hub and shroud) components. The resulting flow field from the interaction of these components is complex as pressure and velocity change rapidly. Flow field in the guide vane passage is determined by the instantaneous position of the blade relative to the guide vane. Similarly, the flow field in the blade passage is determined by the guide vane alignment relative to the blade. The turbine includes 6 blades and 28 guide vanes; therefore, at any instant of time, the one blade passage interacts with the four guide vanes. As blade approaches the guide vane, both pressure and velocity increase, resulting in very high pressure and velocity locally. Furthermore, wake behind the guide vane trailing edge and the stagnation point at the blade leading edge play a critical role. The combined effect creates the dynamic flow field at the runner inlet, hereafter referred to as a vaneless space. **Figure 5** shows an instantaneous pressure field in the guide vane passage and the runner passage. The contours correspond to the runner pitch ($s$) of 0 - 0.22 at mid-span ($z^* = 0$) of the guide vane and the runner passages. We can see rotor-stator interaction (blade and guide vane) around $s = 0.16$ (marked as Interaction) in the vaneless space. The pressure is high at that location; stagnation point on the blade leading edge—pressure side—interacts with the nearest trailing edge of the guide vane. Pressure at other locations gradually decreases, as distance from the blade increases. Note that the entire pressure field rotates with the runner rotations, and results in a periodic flow phenomenon. The period represents the frequency of rotor-stator interaction. Pressure variation along the runner circumference is presented in **Figure 6**. The locations $R_1$, $R_2$, and $R_3$ correspond to the polylines ($\theta = 0 – 180°$ or $s = 0 – 0.5$) in the vaneless space. The polylines $R_1$ and $R_3$ are near the runner and the guide vane, respectively. The polyline $R_2$ is situated between $R_1$ and $R_3$ at an equidistance. Peak pressure value indicates the flow field around the guide vane trailing edge and the blade leading edge. The pressure gradually reduces along the pitch ($s$) additionally, there are small peaks indicating the flow field at the neighbouring four guide vanes' trailing edge. Large peaks indicate the runner pressure field from the runner blade and small peaks indicate the pressure field from guide vane-to-guide vane. Pressure variation in radial direction, while comparing $R_1$ to $R_2$ to $R_3$, is almost similar and no large gradient is observed.

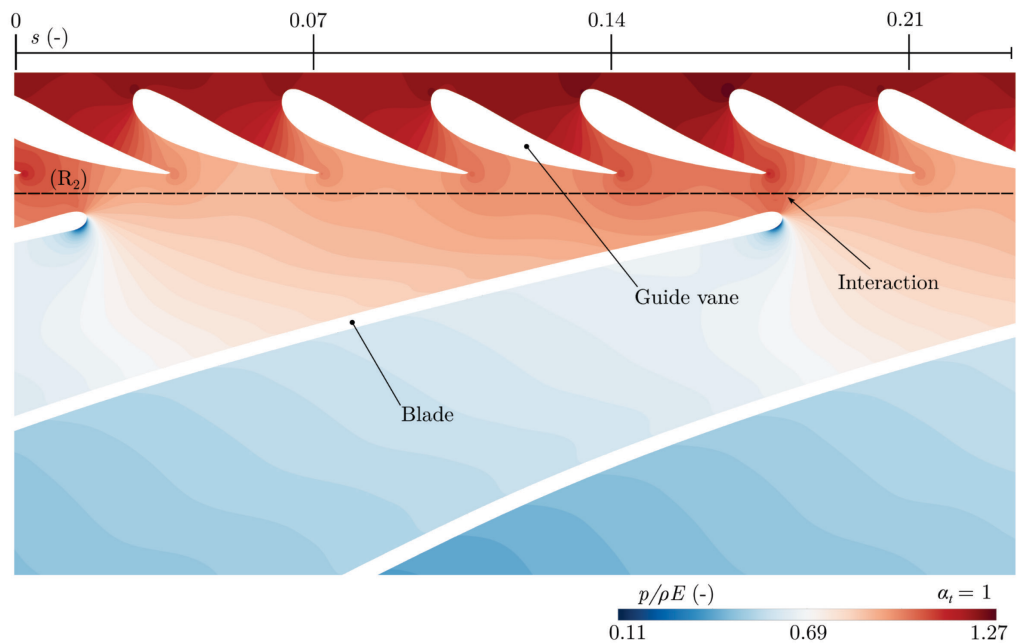

**Figure 5** Contours of pressure loading at mid-span ($z^* = 0$) of the guide vane and runner blade passage.

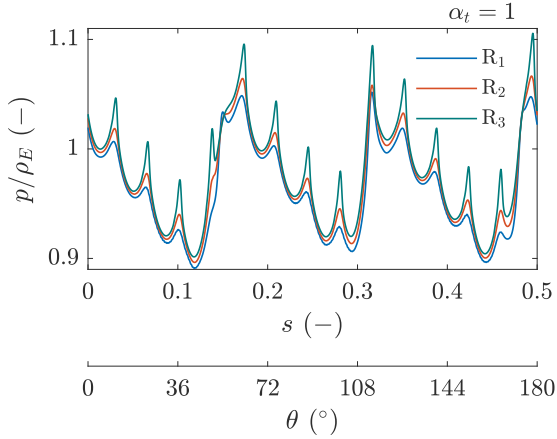

**Figure 6** Pressure variation at the runner inlet in the vaneless space. Pitch ($s$) on x-axis indicates the runner circumference, 0 to 180 degrees.

**Figure 7** shows the radial and tangential velocity along polyline R$_2$ in the vaneless space. The velocity values are normalized with the characteristic velocity ($v_c$). Overall variation in the radial velocity reflects the rotor-stator interaction. The peak values correspond to the blade and guide vane locations. We can see small variations in peak values for s = 0.5 – 1. This is because of the runner blade is not perfectly aligned with the guide vane trailing edge. The tangential velocity shows a clear signature of rotor-stator interaction. Contrary to the radial velocity, we do not see a clear peak of tangential velocity, indicating no significant change in the tangential velocity across the vaneless space. The periodic variation of small peaks around $v_t^* = -0.5$ indicates the influence of the wake from the guide vane trailing edge. The low value ($v_t^* = -0.7$) represents the middle of the blade passage. Overall signature of the pressure, radial velocity, and tangential velocity remain similar, except magnitude, for $α_t = 0.64$ and $α_t = 1.18$ load conditions of the turbine.

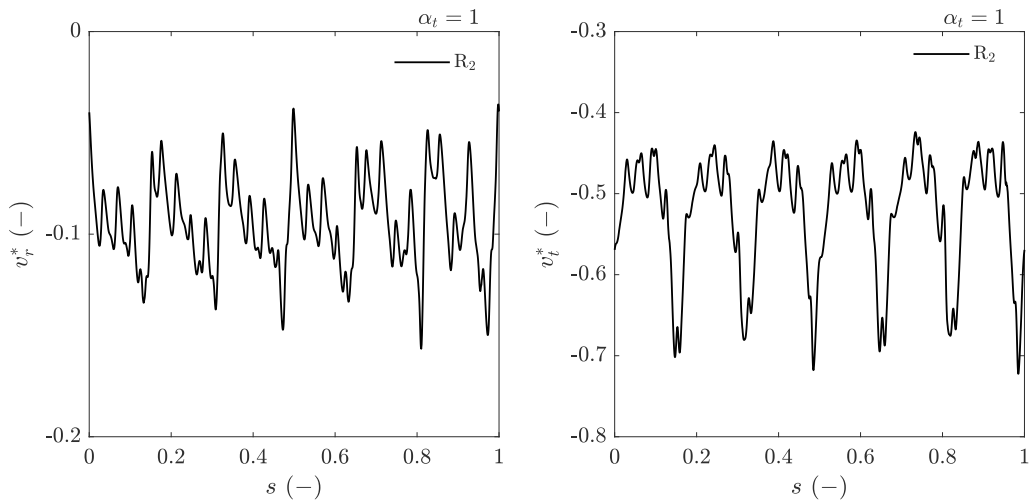

**Figure 7** Radial ($v_r^*$) and tangential ($v_t^*$) velocity at the runner inlet in the vaneless space. Pitch ($s$) on x-axis indicates the runner circumference, 0 to 360 degrees.

$$v_r^* = \frac{v_r}{\sqrt{2gH}} \quad (10)$$

$$v_t^* = \frac{v_t}{\sqrt{2gH}} \quad (11)$$

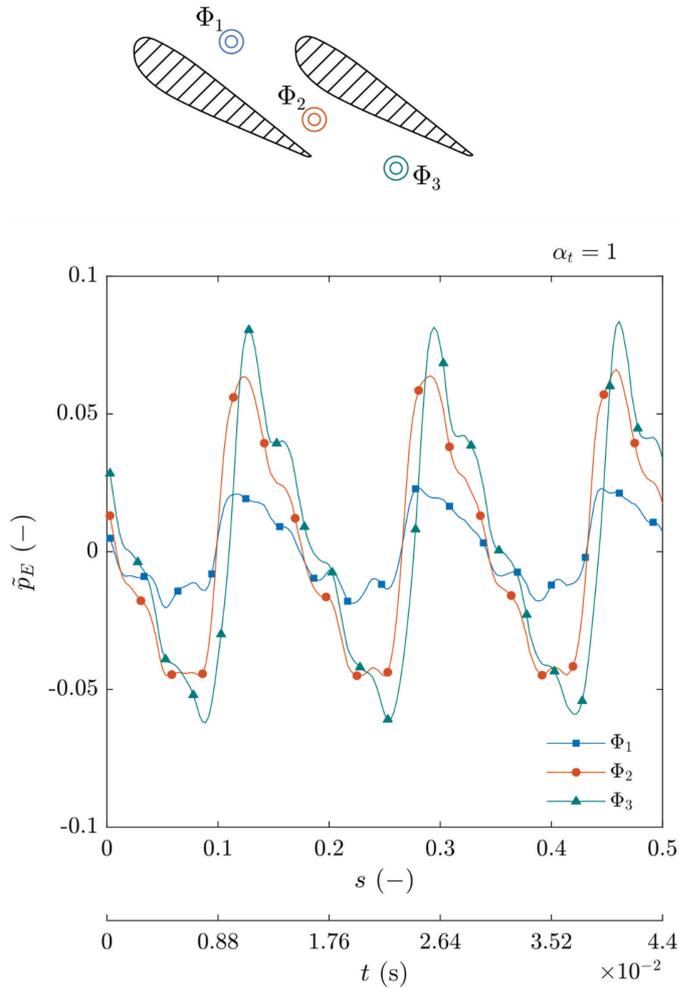

**Figure 8** Time-dependent pressure variation in guide vane passage at design load. Pitch (*s*) on x-axis indicates the runner angular rotation in time. Secondary x-axis indicates the time for 180 degrees runner rotation.

To acquire unsteady pressure data at every time step, with respect to runner angular rotation, several data observation points were created at the start of the simulations. This allowed us to investigate the unsteady flow characteristics with respect to time—as runner rotates. **Figure 8** shows the time-dependent variation of pressure at three locations in the guide vane passage. The locations $\Phi_1$, $\Phi_2$, and $\Phi_3$, show the pressure variation at the guide vane inlet between two guide vanes and the vaneless space, respectively. The pressure values are normalized by specific hydraulic energy ($\rho E$), see **Equation 12**. The pressure values are shown for the pitch, $s = 0.5$, i.e., 180° rotation of the runner in time; $s = 1$ indicates one complete revolution of the runner. The unsteady simulations are conducted for 10 rotations of the runner, and the last five rotations showed stable and converged flow parameters. The presented pressure values are extracted from the last rotation of the runner. The unsteady pressure change corresponds to the blade rotation. High pressure value indicates the blade passes through the nearest observation point. The locations $\Phi_1$, $\Phi_2$, and $\Phi_3$ are aligned in the guide vane passage, and the pressure increases at almost the same time of instant. Peak pressure value indicates the blade is nearest to the location in the vaneless space, i.e., $\Phi_3$. The signature of pressure variation is interesting. When the blade approaches the observation location, e.g., $\Phi_3$, the pressure in the vaneless space rises quickly (see $s = 0.1, 0.29$ and $0.45$), building steep uphill gradient. Once the blade passes the location, the pressure gradually reduces (downhill) to minimum indicating middle of the blade passage, i.e., Two blades are equidistant from the observation location. Small oscillations at the downhill indicate the combined effect of rotor-stator interaction. The frequency of pressure variation corresponds to the blade passing frequency ($f_b$).

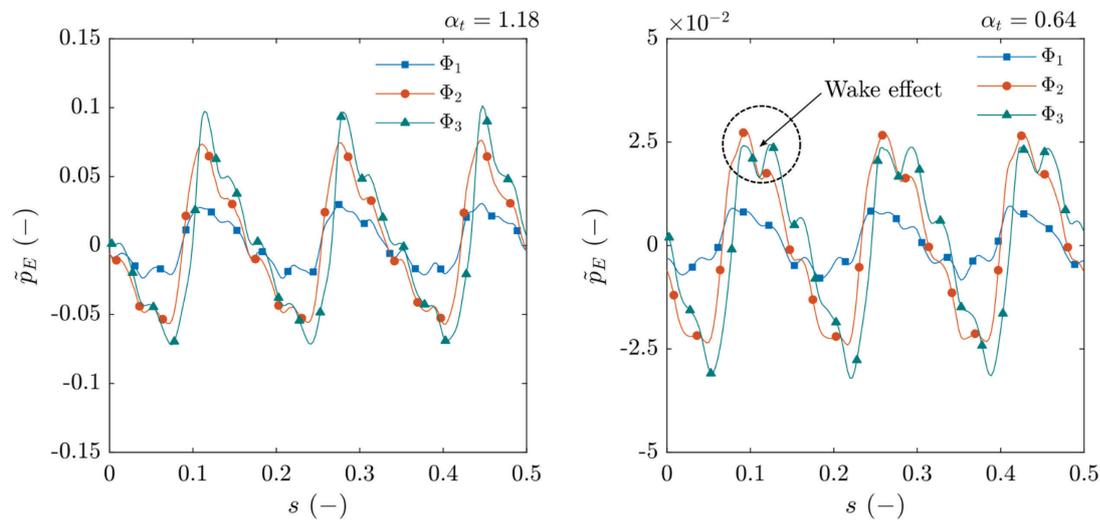

**Figure 9** Time-dependent pressure variation in guide vane passage at part load and high load operating conditions. Pitch ($s$) on x-axis indicates the runner angular rotation in time.

Unsteady pressure variation at high load ($\alpha_t = 1.18$) and part load ($\alpha_t = 0.64$) operating conditions is presented in **Figure 9**. The amplitude of pressure variation for $\alpha_t = 1.18$ is higher than the design load condition. Similarly, at $\alpha_t = 0.64$, the pressure amplitudes are small. Overall signature of the pressure variation appears similar to the design load condition. However, at $\alpha_t = 0.64$, the uphill pressure gradient is not very steep and the pressure in the guide vane passage and the vaneless space is almost equalized. At $\alpha_t = 0.64$, the guide vane angle is small leaving allowing more vaneless space (radial), enabling the uniform pressure field across the guide vanes. A small change in peak pressure value (marked as a dot line circle) at $\Phi_2$ and $\Phi_3$ locations was observed. The variation is attributed to the trailing edge wake. Change in guide vane angle from design load to the part load resulted in alignment of the observation locations on the path of the trailing edge wake.

C Trivedi

$$\tilde{p}_E = \frac{p' - \bar{p}}{\rho E} \tag{12}$$

$$f_b = n \cdot Z_b \tag{13}$$

Spectral analysis has been conducted to study potential frequencies and the amplitudes of the pressure fluctuations. **Figure 10** shows the frequency spectra of the pressure fluctuations at $\Phi_1$, $\Phi_2$, and $\Phi_3$ locations. The frequency is normalized by the runner angular speed ($n$). Thus, $f/n = 6$ indicates the fundamental frequency of runner blade passing. The fundamental frequency along with up to 6 harmonics ($f/n = 36$) is acquired at all three locations. Amplitude is normalized by the specific hydraulic energy. Amplitudes of the fundamental frequency at $\Phi_1$, $\Phi_2$, and $\Phi_3$ locations are 0.18, 0.47, and 0.55, respectively. The amplitudes of the harmonics are gradually decaying as frequency increases. **Table A1** and **Table A5** show the amplitude of the blade passing frequency and harmonics at $\alpha_t = 0.64$, 1, and 1.18, respectively. The amplitudes at high load operating conditions are maximum as expected. When comparing the fundamental frequency (location $\Phi_3$), the amplitudes at $\alpha_t = 0.64$, 1, and 1.18 are 2.4%, 5.4%, and 6.2% of specific hydraulic energy, respectively. To investigate the rotor-stator interaction at further upstream, i.e., Inlet to the spiral casing, two observation points were created. The observation points $\Lambda_1$ and $\Lambda_2$ are located 3 metres and 0.88 metre from the turbine axis respectively. As these points are in stationary domain upstream of the turbine, the expected frequencies are related to the blade passing frequency. Interestingly, both locations successfully capture the blade passing frequency and the harmonics, although the amplitudes are significantly small. Those are 0.1%, 0.2%, and 0.3% of specific hydraulic energy for $\alpha_t = 0.64$, 1, and 1.18, respectively.

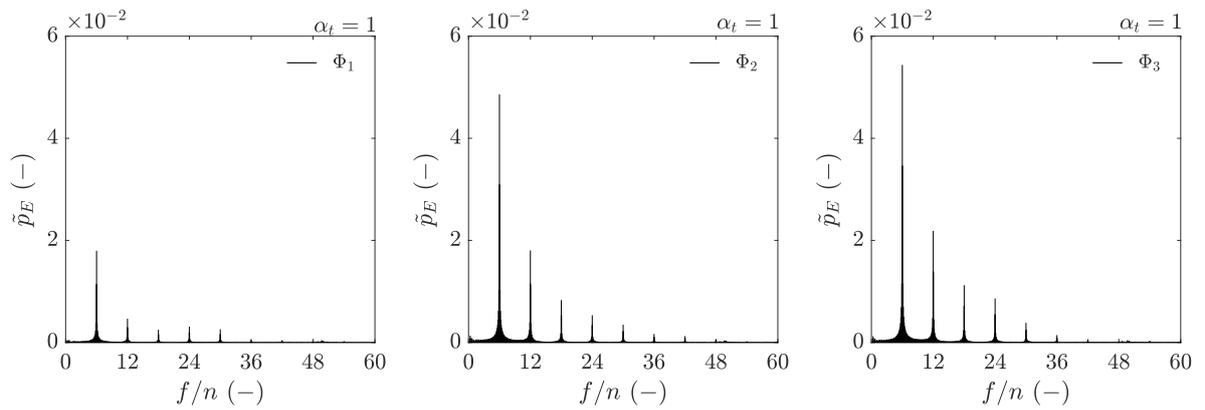

**Figure 10** Frequency spectra of unsteady pressure variation in the guide vane passage at design load. Frequency on x-axis is normalized with the runner angular speed ($n$).

## Blade loading

Runner is a rotating component of the turbine, and the flow field is largely dependent on the incoming flow angle (guide vane angular position), available vaneless space, angular speed of the runner, and blade design. Runner blades extract the potential and kinetic energy of water and deliver lift (mechanical torque). Blade loading shows the energy extractions from the leading edge to the trailing edge. Pump-turbine runner blades are different from that of the Francis turbine. Pump-turbine blades are designed to extract energy and deliver torque in turbine mode, while they use torque and deliver energy (head) in pump mode. **Figure 11** shows blade loading for $\alpha_t = 1$ and 1.18 in turbine mode. The chord ($l/c$) length 0 and 1 represents the leading and trailing edges, respectively. The pressure values are extracted at the mid-span of the blade and normalized by specific hydraulic energy. Large part of the energy is extracted by the first half part of the chord. Overall trend of the pressure loading is almost similar, except for the pressure values for part load, design load, and high load. **Figure 12** shows the contours of the pressure loading on the blades. We can see high and low pressure on the leading edge, pressure side, and suction side, respectively. Pressure distribution across the span from the hub to the shroud is uniform. Overall pressure reduces along the chord of the blade, and we can see low pressure ($p/\rho E = 0.03$) on the trailing edge. **Figure 13** shows the velocity contours at 95% and 5% span of the runner. The span 95% and 5% is near to the hub and shroud of the runner. Flow separation and secondary flow around the leading edge can be seen. The separation results in a small swirling zone near to the leading edge. However, the swirling zone diffuses further downstream in the blade channels. Apart from this, the flow field in the blade channel is nearly stable. The numerical model is also managed to capture the blade trailing edge vortex, which can be seen at both 95% and 5% span locations.

$$v^* = \frac{\sqrt{u^2+v^2+w^2}}{\sqrt{2gH}} \qquad (14)$$

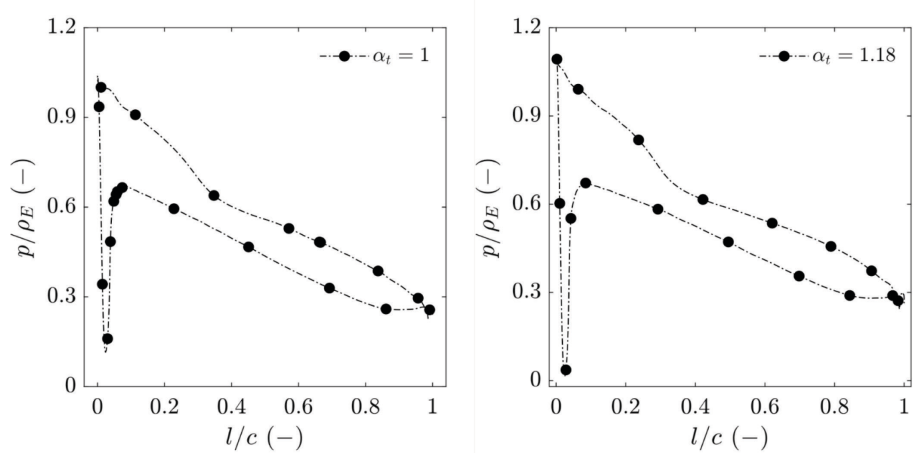

**Figure 11** Pressure loading on the blade at $\alpha_t = 1$ and $\alpha_t = 1.18$. Chord ($l/c$) = 0 refers to the leading edge of the blade.

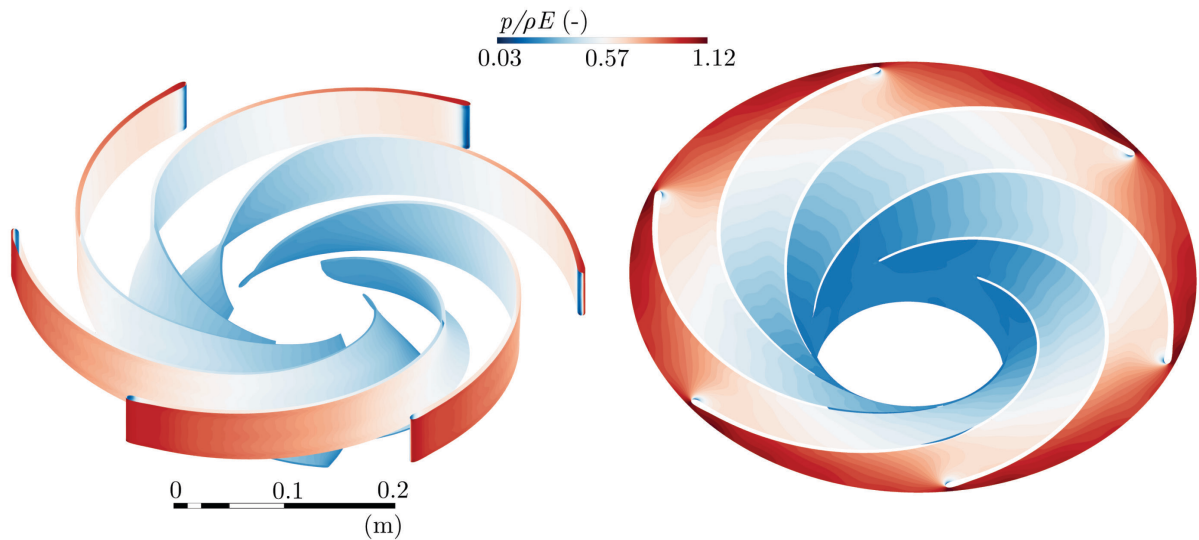

**Figure 12** Contours of pressure loading on the blade at $α_t = 1$.

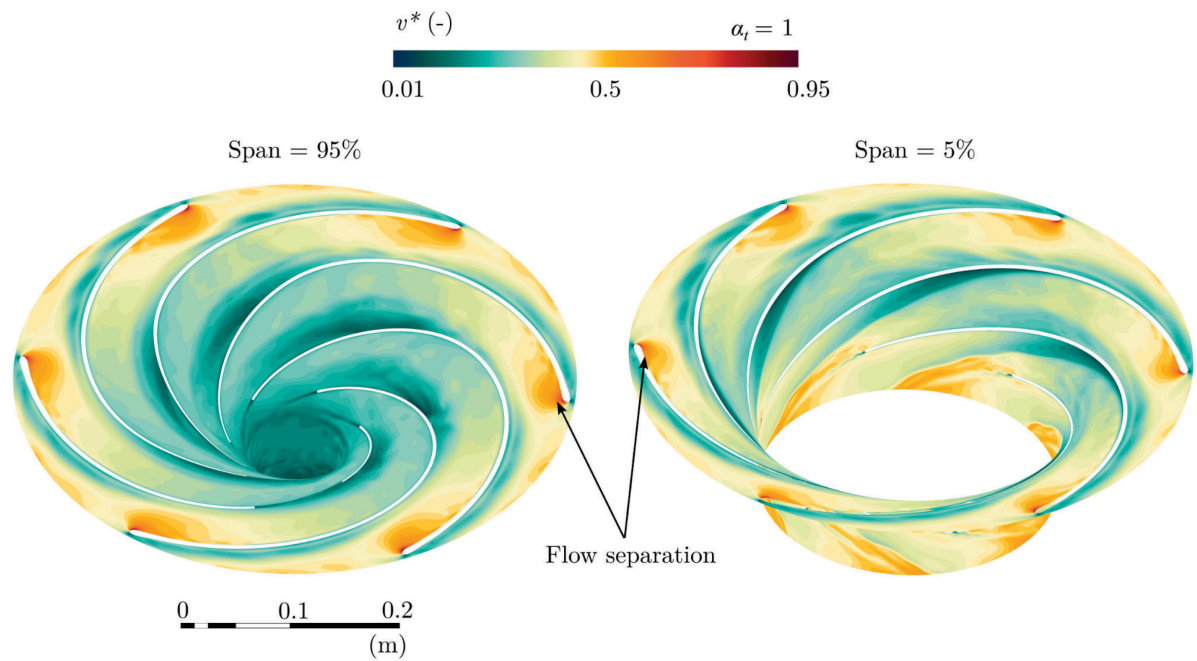

**Figure 13** Contours of velocity at 95% (near to hub) and 5% (near to shroud) span of the runner at $α_t = 1$.

Further analysis is carried out in time-domain to investigate the pressure fluctuations with respect to the blade movement. Total eight observation points were created around the blade mid-span. Points $Π_1$ and $Π_2$ are placed at leading edge and trailing edge, respectively. Points $Π_3$, $Π_4$ and $Π_5$ are placed at 25%, 50%, and 75% of the chord of the blade on the pressure side, respectively. Points $Π_6$, $Π_7$ and $Π_8$ are placed at 25%, 50%, and 75% chord on the suction side, respectively. **Figure 14** shows the unsteady pressure fluctuations at the leading edge and the trailing edge of the blade. High amplitude fluctuations represent the effect of guide vanes on the blade leading edge. Since observation points are in a rotating domain, they represent the interaction with the stationary domain, guide vane. The frequency of fluctuations corresponds to the guide vane passing frequency ($f_{gv}$). Interestingly, the observation point $\Pi_1$ also acquired the fundamental frequency of runner rotation, indicating the rotation of the pressure field with the runner. For example, around $s = 0.25$, the amplitude of pressure fluctuations is high ($\tilde{p}_E = 0.11$) and gradually reduces. Around $s = 1.25$, the amplitude is low ($\tilde{p}_E = 0.05$) and rises quickly to the level $\tilde{p}_E = 0.11$ and cycle repeats. The point $\Pi_2$ is exactly on the trailing edge and shows the unsteady effect of the blade trailing edge wake. The amplitude is significantly small, and no specific signature of fluctuations is observed.

$$f_{gv} = n \cdot Z_{gv} \tag{15}$$

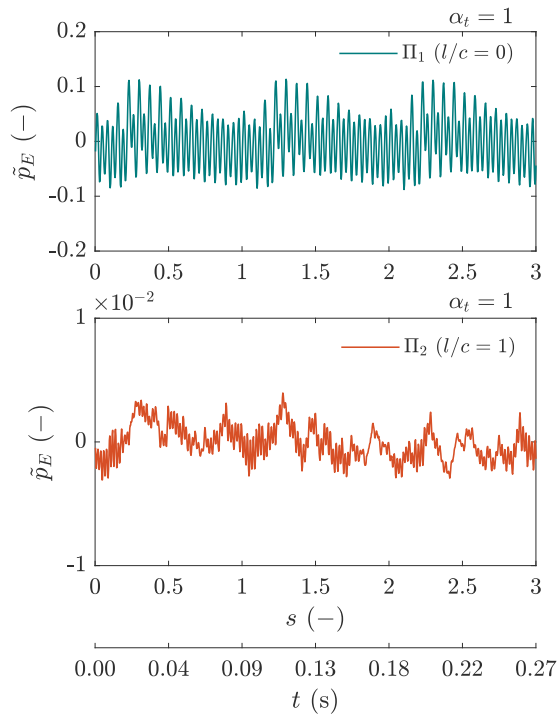

**Figure 14** Time-dependent pressure fluctuations on the leading edge and trailing edge of the blade. Pitch (*s*) on x-axis indicates the runner rotation, and pressure variation for three revolutions is shown. The secondary x-axis shows the actual time in second. Note: y-axis scale for the leading edge and trailing edge points is different.

Unsteady pressure variation on the pressure side and suction side points is shown in **Figure 15**. The fluctuations are shown for 0.2 revolutions of the runner. The amplitude of pressure fluctuations at location $\Pi_3$ is high, around 0.02, and at further down to the blade chord, the amplitude reduces significantly, around 0.001 for $\Pi_5$. Interestingly, on the blade suction side locations, $\Pi_6$, $\Pi_7$, and $\Pi_8$, the amplitudes are less than 50% of that of the pressure side, and they are almost similar in all locations, indicating the effect of the guide vane passing frequency similar across the blade. Strong flow separation and the local swirling flow near to the suction side result in lower pressure on the blade suction side. Unsteady pressure fluctuations at part load ($\alpha_t = 0.64$) and high load ($\alpha_t = 1.18$) operating conditions are shown in **Figure 16**. Overall signature of the pressure fluctuations is identical. The pressure loading for $\Pi_3$ is 0.01 and 0.02 at part load and high load, respectively. For the points $\Pi_4$ and $\Pi_5$, no significant difference is obtained, indicating the blade loading for chord length $0.5 - 1$ is nearly the same for the stable operating conditions of the turbine.

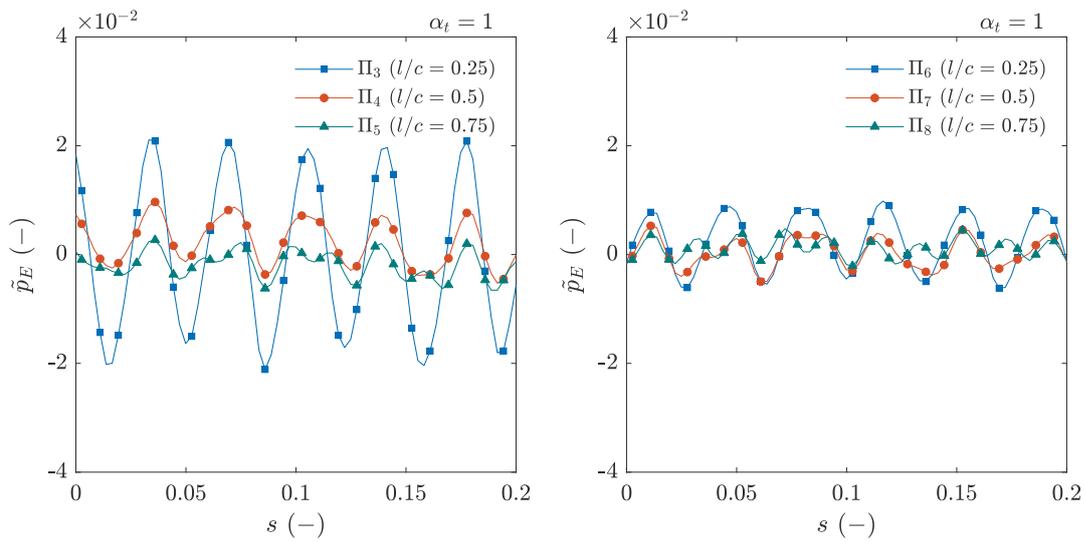

**Figure 15** Time-dependent pressure on the pressure side and suction side of the blade at different chord length in mid-span. Pitch ($s$) on x-axis indicates the runner rotation, and pressure variation for 72 degrees revolution is shown.

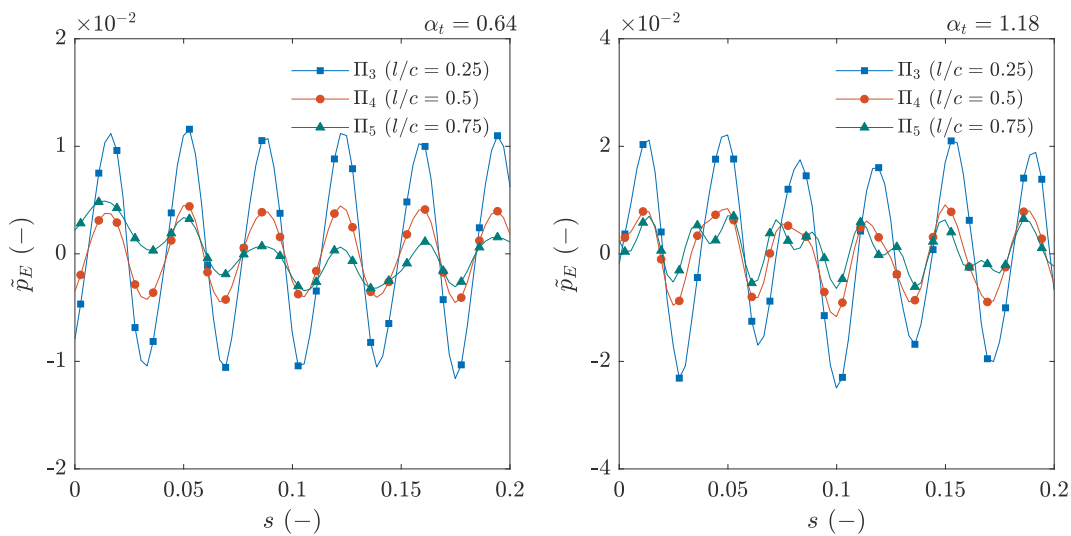

**Figure 16** Time-dependent pressure on the pressure side and suction side of the blade part load ($\alpha_t = 0.64$) and high load ($\alpha_t = 1.18$) operating conditions. Pitch ($s$) on x-axis indicates the runner rotation, and pressure variation for 72 degrees revolution is shown. Note: y-axis scale for the part load and high load is different.

C Trivedi

Spectral analysis is conducted to investigate the specific frequencies of the unsteady pressure fluctuations. Extracted frequency spectra at design load ($\alpha_t = 1$) is shown in **Figure 17**. The frequency and amplitude are shown for three locations: blade leading edge, pressure side (25% chord), and suction side (25% chord). The frequency is normalized by the rotational speed of the runner. The frequency spectra clearly show the frequency of rotor-stator interaction, i.e., guide vane passing, in the runner at all locations. At leading edge location ($\Pi_1$), shows the half of the fundamental frequency ($f/n = 14$), the fundamental frequency ($f/n = 28$), and the second harmonic ($f/n = 42$). As discussed previously, the frequency of runner rotation ($f/n = 1$) is also acquired, and several harmonics of the runner rotation can be seen. The fundamental frequency ($f/n = 1$) has the maximum amplitude. When investigating the frequency on the blade pressure side and suctions side, locations $\Pi_3$ and $\Pi_6$, the predominant frequency corresponds to the guide vane passing frequency, and no other frequency was obtained. Detailed amplitudes and frequencies are shown in **Appendix A**. We can see an interesting signature of the rotor-stator interaction when comparing it to the different load conditions and the different locations. The amplitude (location $\Pi_1$) of fundamental frequency and the half of the fundamental frequency at part load ($\alpha_t = 0.64$) is more than double when comparing to the design load ($\alpha_t = 1$). Furthermore, at some of the locations, the half of the fundamental frequency and the second harmonic are not available. At high load ($\alpha_t = 1.18$) condition, all three frequencies are acquired for all observation locations, although the amplitude at downstream ($l/c > 0.5$) is not high ($\tilde{p}_E < 1\%$).

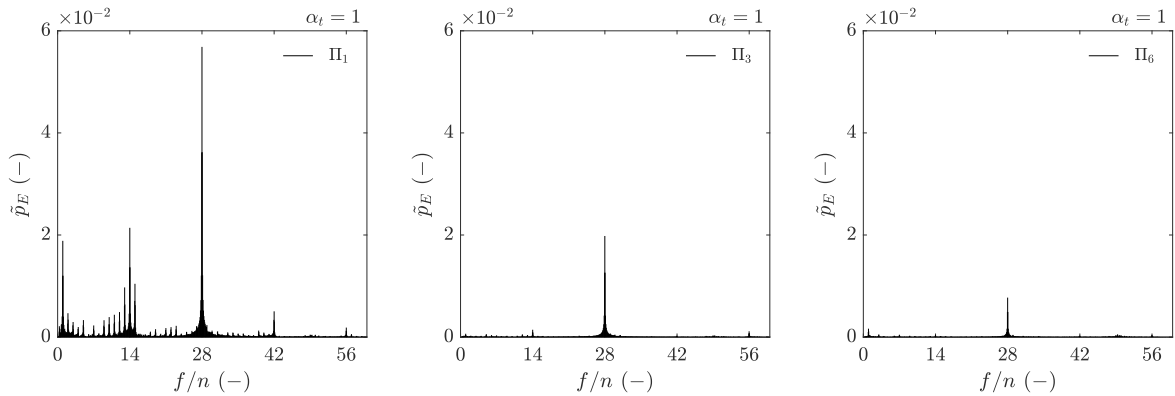

**Figure 17** Frequency spectra of the unsteady pressure fluctuations at three locations of the runner blade. The frequency is normalized by the runner rotational speed.

## Interaction of blade and draft tube

Flow in the draft tube is largely dependent on the flow leaving the blade channel and the trailing edge vortex in the turbine mode operation. Furthermore, main function of the draft tube is recovering the pressure energy. Therefore, is generally designed as a diffuser, where cross-section area is increasing along the flow path. This leads to adverse gradient of pressure near the no-slip wall of the draft tube.

This results in combined effect of runner flow leaving the runner and the secondary flow due to separation from the draft wall. To investigate the flow leaving the runner, four observation points are created at the edge of the runner outlet and draft tube inlet. The observation points $\psi_1$, $\psi_2$, $\psi_3$ and $\psi_4$ are located at $z^* = 1.07$ and. The four points are 90 degrees apart from each other on the same radial plane ($r^* = 1$), resulting in the points $\psi_1$, and $\psi_3$ are opposite (180 degrees) to each other. **Figure 18** shows the unsteady pressure variation at these four locations. The pressure variation is presented for one complete rotation of the runner. The peak values indicate the runner blade position near to the observation location. The frequency of the period is the blade passing frequency ($f_b$). Although these four points are placed on the same plane, the signature of pressure fluctuations varies significantly. Location $\psi_1$ shows a large period of fluctuations with second harmonic at peak pressure value. The signal of continuous sinusoidal wave. Location $\psi_2$ shows the steep gradient of fluctuations at peak value, and the large frequency oscillations at the low-pressure side, i.e., valley point. Location $\psi_3$ shows similar signature of steep gradient as $\psi_2$. Location $\psi_4$ shows completely different signature with steep valley and peaks of several high frequencies. Although these observations points are on the same radial plane, their signature of pressure fluctuations is largely dependent on the spatial locations at the runner downstream.

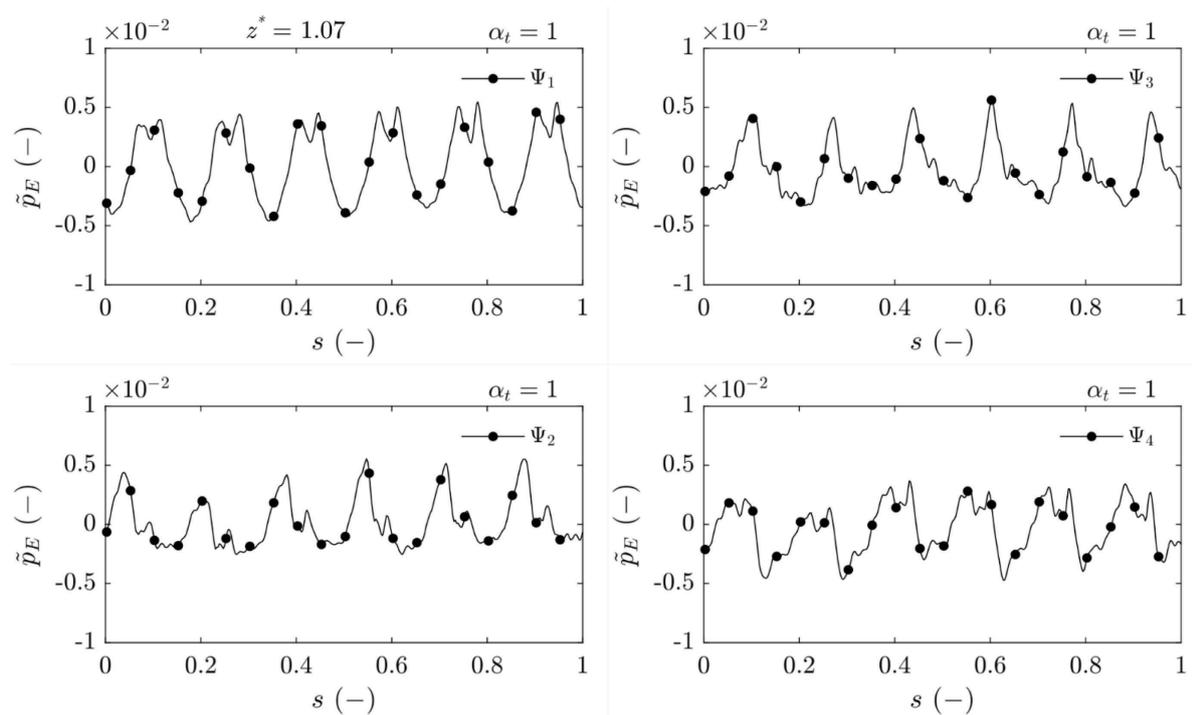

**Figure 18** Unsteady pressure variation at the runner outlet ($z^* = 1.07$) in the draft tube at design load.

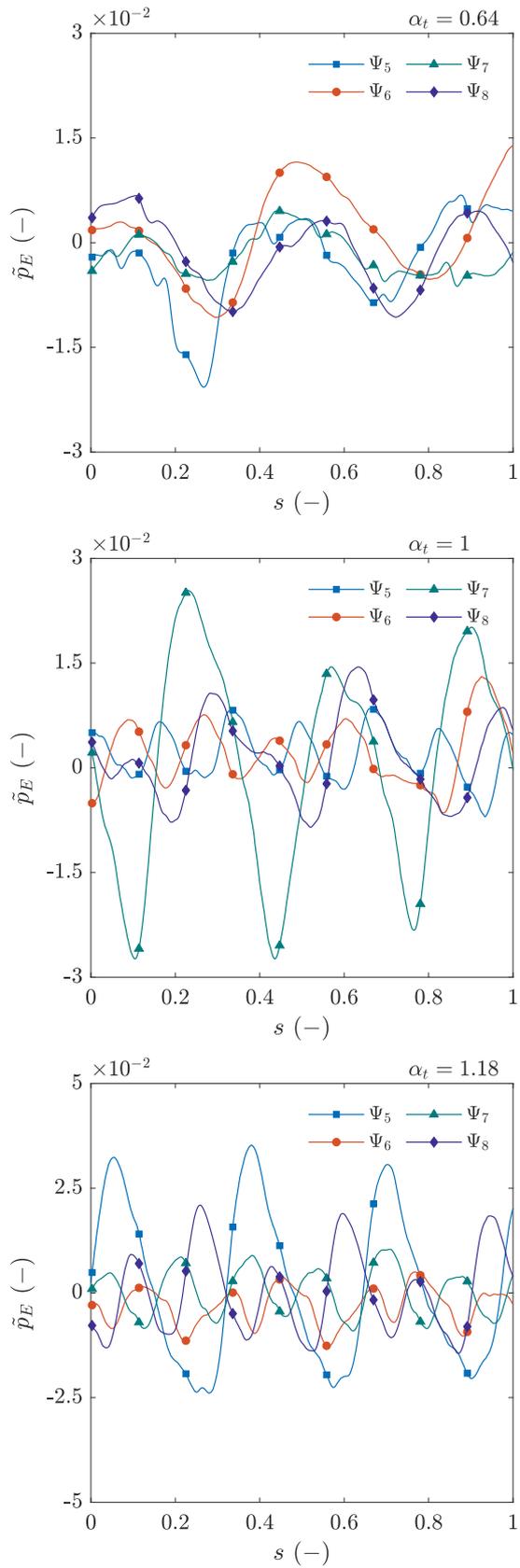

**Figure 19** Unsteady pressure variation at the runner outlet ($z^* = 1.07$) in the draft tube at part load, design load and high load.

Unsteady pressure fluctuations at another axial distance ($z^* = 1.75$) in the draft tube are shown in **Figure 19**. The comparison of pressure fluctuations for different load conditions is presented. Observation points ($\psi_5$, $\psi_6$, $\psi_7$, and $\psi_8$) are 90 degrees apart from each other on the same radii ($r^* = 1.03$). At part load ($\alpha_t = 0.64$), we can see a period of low frequency, and it is in-phase. We do not see a clear signature of the vortex rope, although there is a small phase shift. The period shows the combined flow field of the runner's angular rotation and the vortex rope. At design load ($\alpha_t = 1$), we can see significant variations in the fluctuations. Fluctuations at the location $\psi_7$ have more than double the amplitude of the other locations on the same radii. At high load ($\alpha_t = 1.18$), we can see a similar variation; however, the high amplitude fluctuations correspond to the location $\psi_5$. The location $\psi_8$ also shows somewhat high amplitude. Spectral analysis of the observation points $\psi_5$, $\psi_7$, and $\psi_5$ at part load, design load, and high load conditions is shown in **Figure 20**. At part load (location $\psi_5$), several low frequency pressure fluctuations are present; however, the predominant frequencies are related to the runner rotation and harmonic. Furthermore, the frequency of runner blade passing does not exist. At design load (location $\psi_7$), three frequencies are obtained. The first visible frequency is $f/n = 0.3$, which appears to be the vortex rope frequency; however, the amplitude is very small ($\tilde{p}_E = 0.003$). The second frequency is predominant, and it is the half of the fundamental frequency ($f/n = 3$) of the runner blade passing. The third frequency is the fundamental frequency ($f/n = 6$) of the runner blade passing. Spectral analysis at high load location $\psi_5$ shows three specific frequencies of high amplitude. Those frequencies correspond to the half of the fundamental frequency of the runner blade passing, the fundamental frequency, and the second harmonic of the runner blade passing.

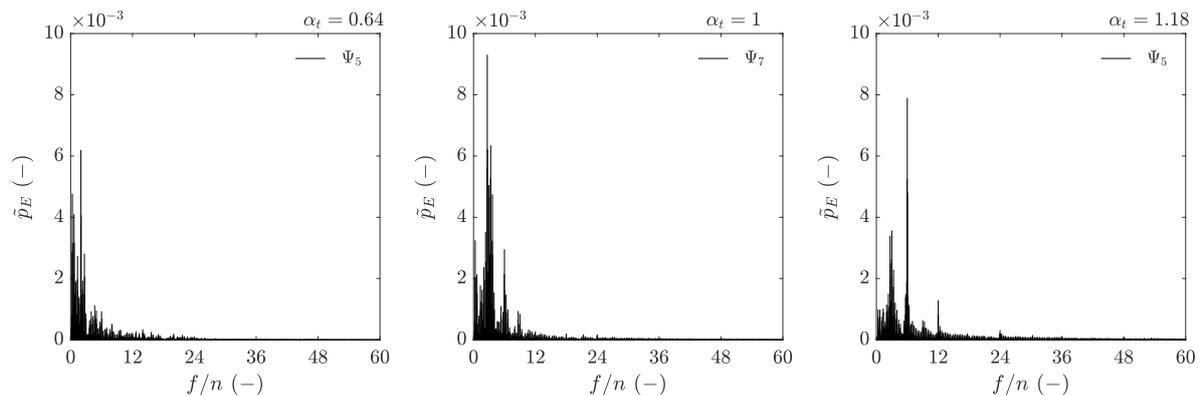

**Figure 20** Frequency spectra of the unsteady pressure fluctuations in the draft tube ($z^* = 1.75$, $r^* = 1.03$) at three operating conditions of the turbine. The frequency is normalized by the runner rotational speed.

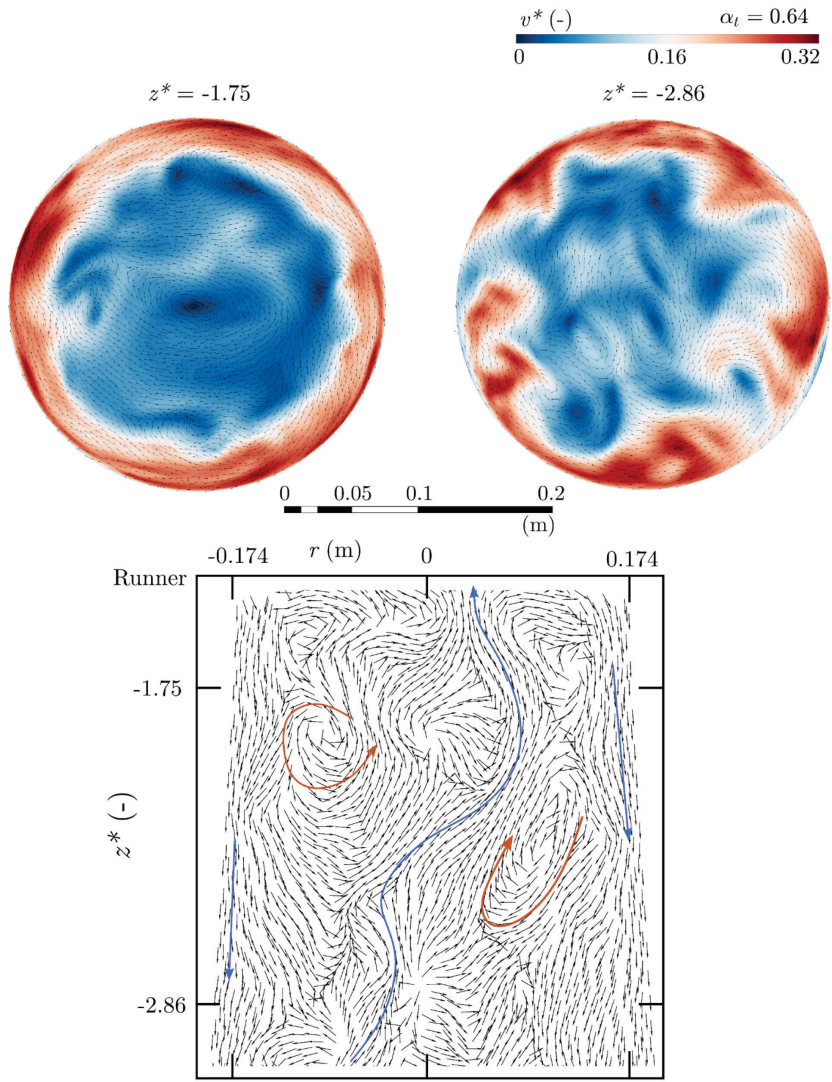

**Figure 21** Velocity contours and vectors at $z^* = 1.75$ and $z^* = 2.86$ planes in the draft tube.

**Appendix A** summarises the amplitude of the blade passing frequency and the harmonics at two locations $\psi_1$, $\psi_5$. To study draft tube flow, velocity contours and the vectors are analysed and presented in **Figure 21**. The flow condition corresponds to the part load, and the contours are shown for two axial planes, $z^* = 1.85$ and $z^* = 2.86$. The bulk rotation can be seen near the wall of the draft tube. Several local swirls are present in the core region, and the rotational direction is similar to the runner. At further downstream ($z^* = 2.86$), the flow is highly stochastic and localised. Several localised swirling zones can be seen, and they have local flow directions, both clockwise and counterclockwise. Vectors on the z-x plane also show highly stochastic flow in axial direction from runner outlet to the downstream below $z^* = 2.86$. Flow near the draft tube wall is accelerating downward. However, in the core region, the flow appears reversible up to the runner and splits into two regions. In addition, several local swirls are clearly visible, marked by arrows.

# Pump mode

## Secondary flow in draft tube

Numerical investigations are conducted at four operating conditions in pump mode operation. The runner rotates in counter direction, and the flow direction is reversed. The flow enters from the draft tube to the runner then guide vanes and leaves from the spiral casing. For the pump mode operation, $\alpha_p = 1$ refers to the reference load condition for pump mode operation. The load at $\alpha_p = 0.64$, $\alpha_p = 1.18$, and $\alpha_p = 1.3$ indicates the part load, high load, and full load conditions, respectively. Key findings from the selected operating conditions are presented in this section although the numerical results are investigated at all operating conditions. The locations and name conventions of the observation points are identical to the turbine mode.

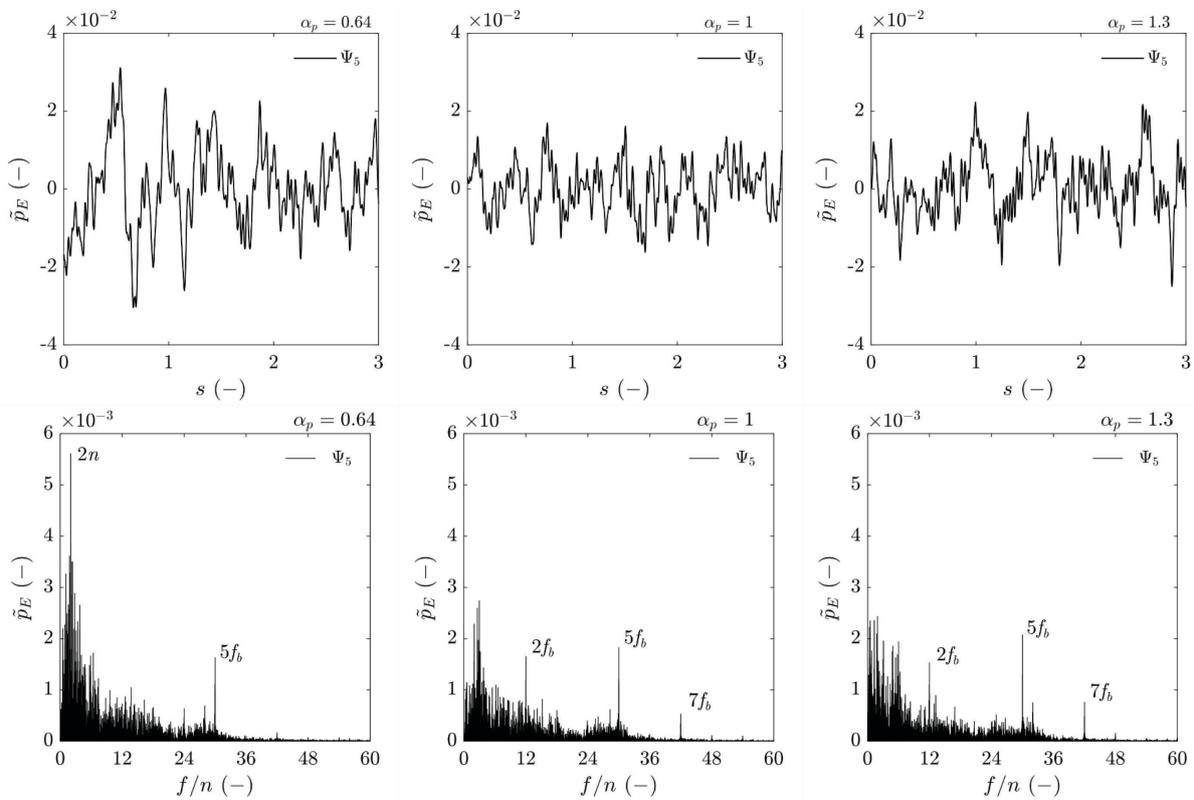

**Figure 22** Unsteady pressure fluctuations and the frequency spectra in the draft tube ($z^* = 1.75$, $r^* = 1.03$) for part load, design load and full load conditions.

For pump mode, the inlet component of the turbine is the draft tube, and the investigated turbine is equipped with an elbow type draft tube. Flow field in the draft tube is largely dependent on the geometrical changes along the flow path. Flow from the inlet of the draft tube is streamlined as cross-section area reduces gradually up to the elbow. From the elbow, the draft tube is aligned with the runner axis, and the cross-section area reduces further to match the runner diameter. There are four observation points ($\psi_5$, $\psi_6$, $\psi_7$, and $\psi_8$) in the draft tube at $z^* = 1.75$, which is around 0.125 m from the runner and draft tube interface. Unsteady pressure fluctuations and the frequency spectra at one of the locations ($\psi_5$) are presented in **Figure 22**. The pressure fluctuations and the frequency spectra are shown for part load, design load, and full load conditions. The fluctuations are largely stochastic type indicating the presence of the secondary flow in the draft tube. The spectral analysis reveals wide band frequency related to the stochastic fluctuations that can be seen up to $f/n = 36$. Interestingly, the spectral analysis also shows the frequency associated with the runner rotational speed ($f/n = 1$) and the harmonics that clearly indicate the runner flow field interaction with the draft tube incoming flow. The frequency of blade passing ($f_b$) is also observed with small amplitude for all operating points.

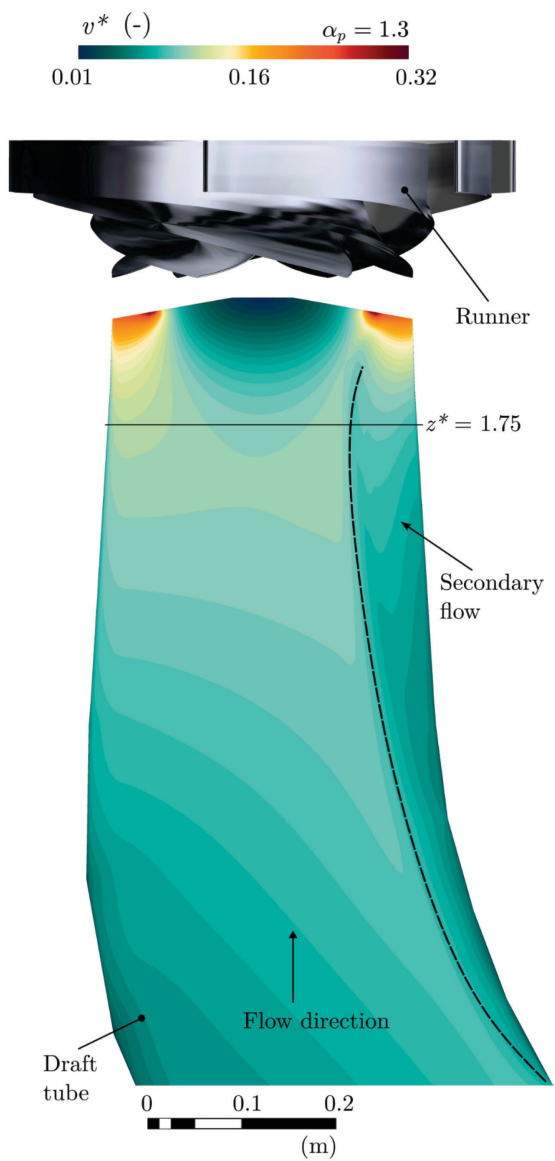

**Figure 23** Contours of velocity distribution in the draft tube and inception of secondary flow from the elbow, Dean vortices.

Draft tube of the pump-turbine is an elbow type, and the secondary flow from the 90-degree elbow is obtained, widely referred to as Dean vortices [36]. The presence of Dean vortices is clearly seen in the draft tube; see **Figure 23**. Flow on the small radius bend separates from the runner inlet. The velocity values are normalised with the maximum velocity, **Equation 14**. The separation boundary is marked with a dash line from the elbow to the runner inlet. Change in the velocity values within and outside the separation zone can be seen. The separation is the result of the radial pressure gradient — high pressure outside and low pressure inside the zone. Similarly, the velocity gradient is also present — high velocity near the wall and low velocity in the centre of the of the draft tube core. Two opposing vortices are formed near to the small radius of the elbow. **Figure 24** shows the contours of velocity and vectors representing the Dean vortices at the plane $z^* = 1.75$. High rotational velocity ($v^* = 0.13$) can be seen near the wall of the draft tube, and the flow moves towards the centre of the draft tube. At the same time, flow from the centre of the draft tube moves outward radii. The flow interacts at around $r^* = 0.97$, marked as a shear line. On the elbow side ($x^* = 1.03$, $y^* = 0$), the inception of the Dean vortices can be seen. The Dean vortices draw momentum from the bulk flow near to the wall and pull inward, rotate in opposite directions. The radial flow from the centre of the draft tube pushed over the edge of the Dean vortices then pulled into the core of the Dean vortices.

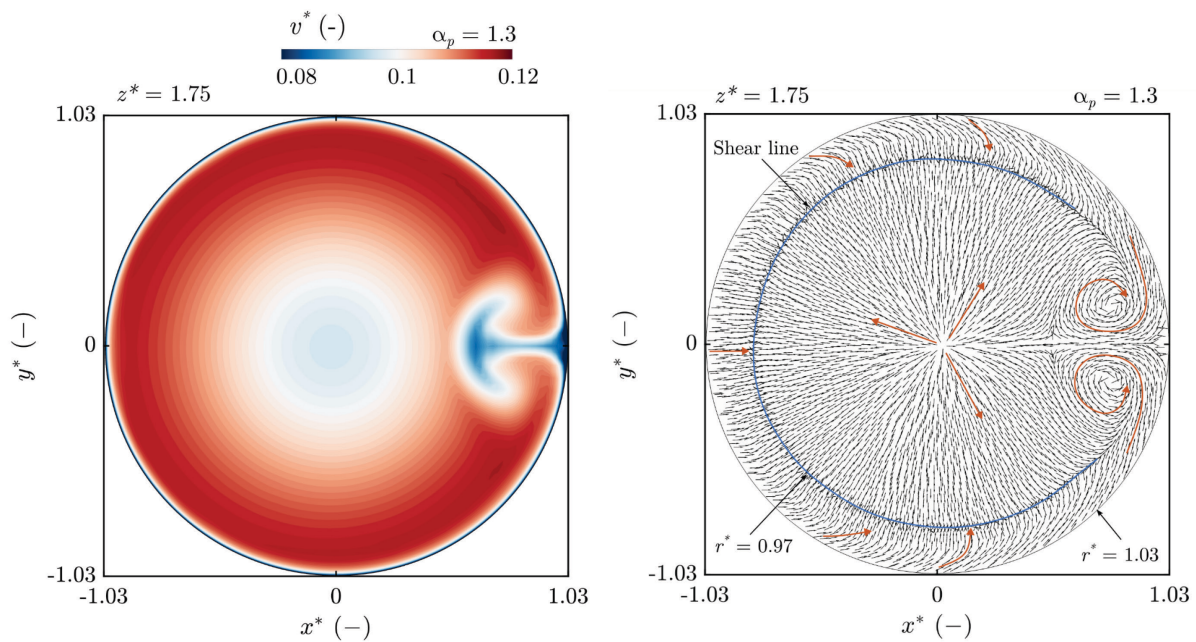

**Figure 24** Velocity contour and vector representing the Dean vortices in draft tube cone at $z^* = 1.75$. Dimensions on x-axis and y-axis are normalized by the reference radius of 0.1754 m.

## Blade loading and passage vortex

Flow to the runner enters from the draft tube in pump mode operation. Therefore, the flow can be symmetric or asymmetric depending on the design of the draft tube. We studied the effect of the draft tube bend and resulting Dean vortices. In the present draft tube, the effect of the Dean vortices is minimum at the runner inlet— runner and draft tube interface. Near to the interface, the effect of the runner blades is predominant. In pump mode, the flow starts from the blade trailing edge and leaves

C Trivedi

the leading edge towards the guide vane. This section describes the results obtained from the runner domain. Note the name conventions follow the turbine mode, e.g., blade leading edge, trailing edge pressure side and suction side. However, the dimensions of the chord are reversed, e.g., $l/c = 0$ is the blade trailing edge and $l/c = 1$ is the blade leading edge. **Figure 25** shows the blade loading at part load ($\alpha_p = 0.64$), design load ($\alpha_p = 1$), and full load ($\alpha_p = 1.3$). The blade loading along the chord increases, from the trailing edge to the leading edge. The maximum pressure is around 1.5 times the specific hydraulic energy ($\rho E$) on the leading edge. **Figure 26** shows contours of the pressure loading on all blades. We do not see significant change in span-wise pressure loading. However, we can see a high-pressure gradient on the blade suction side around the trailing edge. This is the result of local separation of the flow from the trailing edge on the suction side. Negative pressure was obtained at specific locations on the trailing edge and the suction side, indicating the possibility of local cavitation spots. Cavitation is not modelled for the present study.

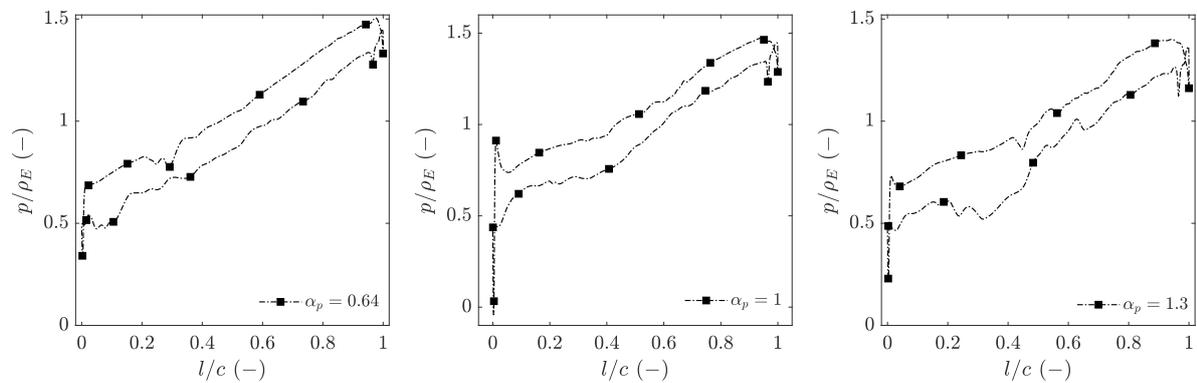

**Figure 25** Blade loading at mid-span of the turbine blade for part load, design load and full load. Chord length ($l/c$) of 0 and 1 corresponds the blade trailing edge and leading edge, respectively.

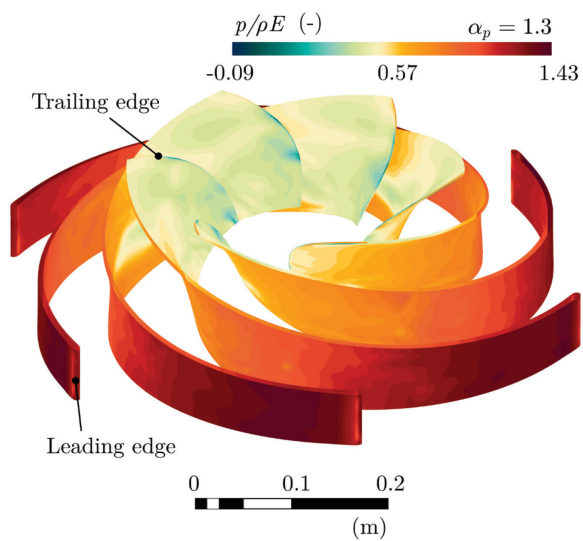

**Figure 26** Contours of blade loading in pump mode operation at full load. The runner is flipped towards the trailing edge to highlight the low pressure on the blade suction side.

C Trivedi

To investigate time-dependent unsteady pressure fluctuations on the blade, eight observation points are created, following the turbine mode operation. The points $\Pi_1$ and $\Pi_2$ are on the blade leading edge and trailing edge, respectively. The points $\Pi_3$, $\Pi_4$, $\Pi_5$ are on the blade pressure side at different chord length. The points $\Pi_6$, $\Pi_7$, $\Pi_8$ are on the blade suction side at different chord length. Unsteady pressure fluctuations at full load condition are shown in **Figure 27**. A large variation in the pressure fluctuations is observed among different locations of the blade. Pressure on and around the blade trailing edge locations ($\Pi_2$, $\Pi_5$ and $\Pi_8$) is asymmetric and highly stochastic. Moreover, the stochastic fluctuations are high amplitude and low frequency. As flow travels towards the outlet, the signature of the fluctuation changes from a highly stochastic to a systematic. The locations $\Pi_4$ and $\Pi_7$ show fluctuations of both low and high frequencies. The systematic fluctuations of high frequency are predominant at the locations $\Pi_3$ and $\Pi_6$, which are associated with the guide vane passing frequency and the runner rotational frequency. Spectral analysis of the pressure fluctuations at three locations is presented in **Figure 28**. The location at the blade trailing edge, $\Pi_2$, shows low frequency of wide band indication predominantly stochastic fluctuations around the trailing edge. Interestingly, the spectra also show the fifth harmonic of the blade passing frequency. It also shows the frequency of runner angular speed with the maximum amplitude. The spectra on $\Pi_1$ leading location clearly show the guide vane passing frequency, including the fifth harmonic of the blade. We can also see the other frequencies of small amplitudes indicating the presence of wake and the vortex shedding from the leading edge. The spectra on $\Pi_3$ location on blade pressure side ($l/c = 0.75$) show a similar signature. The wide band small amplitude frequencies represent the flow unsteadiness and local swirls inside the blade passage. A complete detail of the frequencies and the amplitudes is presented in **Appendix B**.

In pump mode operation, the flow separation from the blade trailing edge suction side towards the blade channel is observed. The swirling flow in the blade passage can be seen in the **Figure 29**. The contour corresponds to the normalised total velocity (see **Equation 14**), mid-span of the blade passage extracted from the last time step of the simulation results. Each passage shows a different flow field depending on the last runner's position relative to the guide vanes. The flow enters from the draft tube side and leaves towards the guide vane side of the runner. A strong trailing edge separation on the suction side can be seen clearly for all blades. A large vortical region is formed that is attached to the suction side of the blade. High velocity from the trailing edge pushes the flow to the blade channel, and the low velocity next to the blade surface moves upward to the trailing edge. The opposite velocity direction and the difference between them create a local swirling zone at the inlet of the passage and rotate in a clockwise direction. Further downstream in the blade passage, another vortical region sets in. However, the vortical region is on the opposite side, near to the pressure side of the neighbouring blade, marked as "Inter-blade vortex" in the **Figure 29**. This swirl spins in a counterclockwise direction. At the outlet of the blade passage, the wake and vortex shedding from the blade leading edge can be seen, marked as "Leading edge wake."

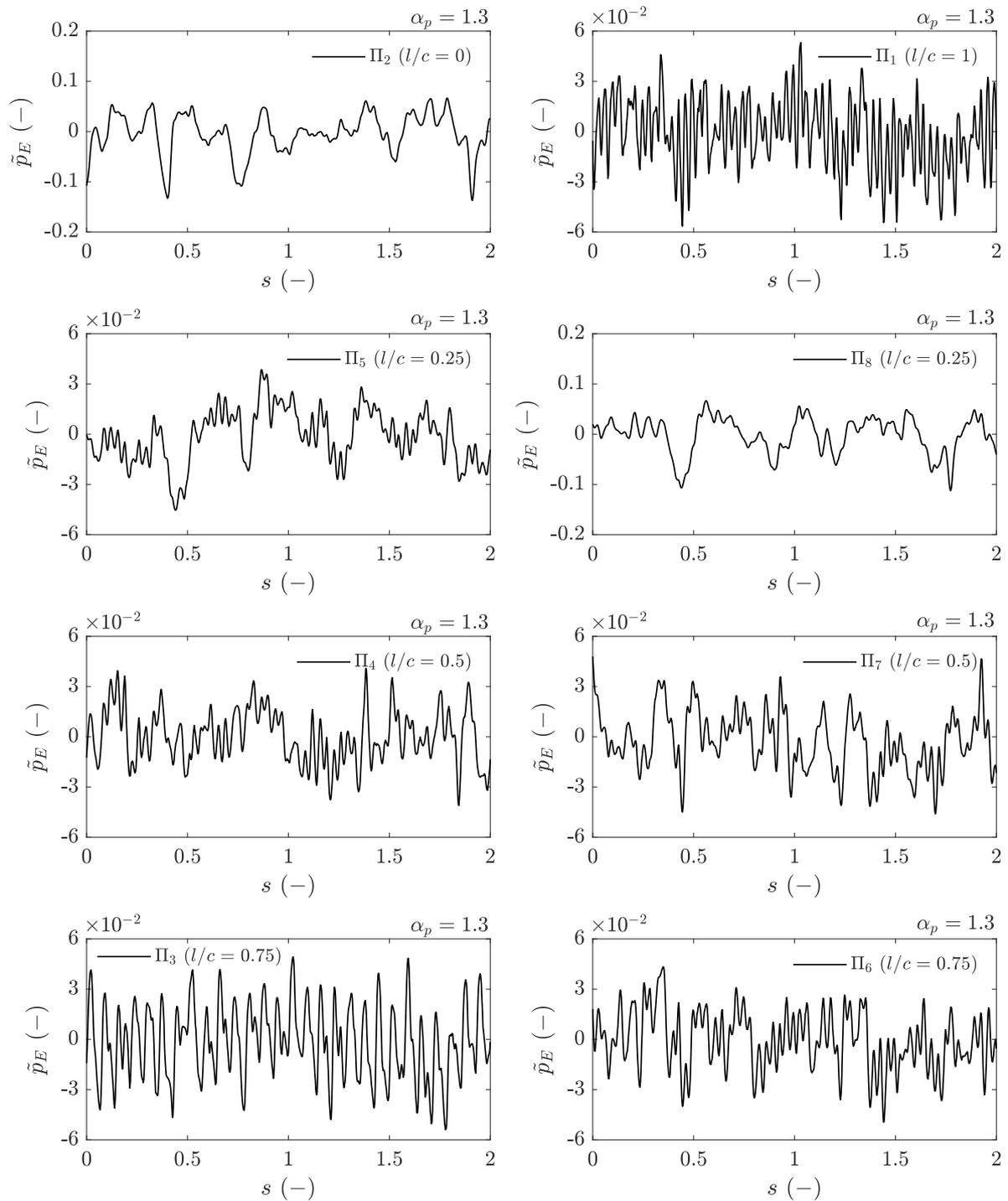

**Figure 27** Unsteady pressure fluctuations on the blade pressure side and suction side at full load. On x-axis, $s = 2$ indicates the two revolutions of the runner. Chord length ($l/c$) of 0 and 1 corresponds the blade trailing edge and leading edge, respectively.

C Trivedi

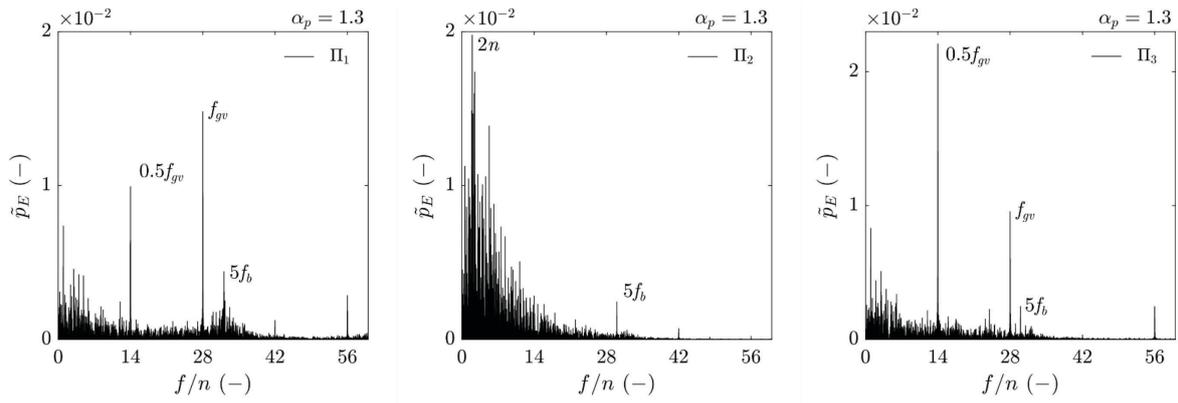

**Figure 28** Frequency spectra of the unsteady pressure fluctuations at three locations of the runner blade in pump mode. The frequency is normalized by the runner rotational speed.

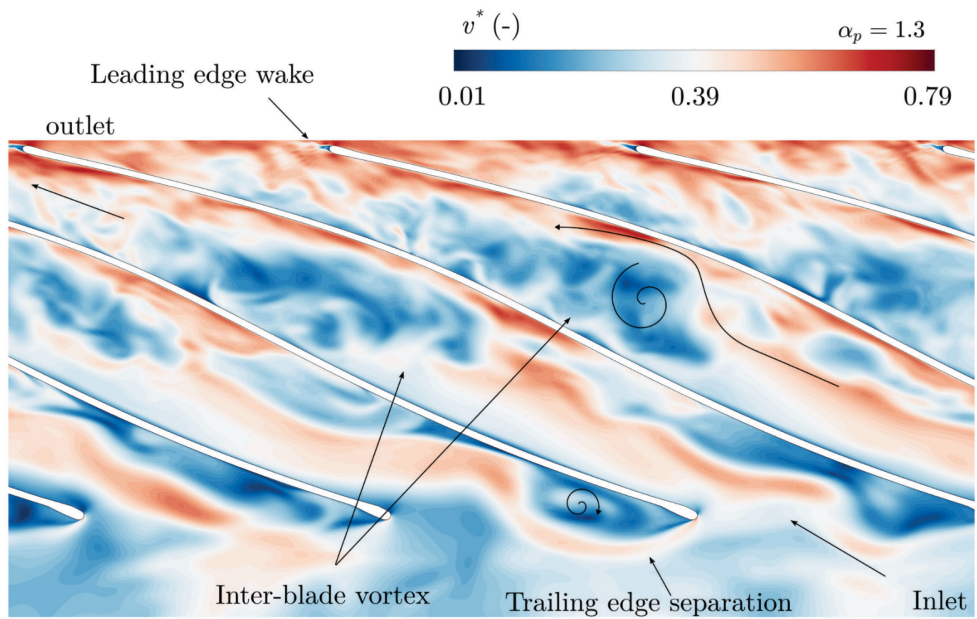

**Figure 29** Contour of total velocity at mid-span of the runner representing the inter-blade vortex at full load.

## Blade leading edge vortex and interaction with guide vane

In a pump mode, flow leaves from the runner and travels through the guide vane and stay vane passage. Unsteady vortex shedding from the blade propagates into the vaneless space; similarly, the vortex shedding from the guide vane leading edge propagates towards the stay vane passage. Rotor-stator interaction in the vaneless space is different from that of the turbine mode operation of the turbine. Flow field is dependent on the rotating blades and rotating wake effect. Change in pressure and the velocity fields in the vaneless space is shown in **Figure 30**. The pressure and velocity data are acquired at the polyline $R_2$ along the runner circumference in the vaneless space. Pitch (*s*) on the x-axis represents an angular position from 0-degree to 180-degree along the runner circumference. Runner is in fixed and same position for all load values. The peak pressure value ($p/\rho E$) indicates the stagnation point of the guide vane trailing edge. When comparing the different flow conditions, no major deviation in the pressure value is obtained. The pressure is low (around the peaks) for full load condition ($\alpha_p = 1.3$). This is related to the guide vane opening position and the available vaneless space. At full load condition, the guide vanes are at the large opening angle, leaving a small vaneless space, resulting in high acceleration of flow through the vaneless space, and the minimum distance between the blade and the guide vane. We can also see changes in pressure gradient — around the valley — when the blade is near to the guide vane, e.g., $s = 0.05$, $s = 0.22$, $s = 0.38$.

The acquired radial and tangential velocity are also shown in **Figure 30**. Both velocities are normalized with the maximum theoretical velocity, **Equation 10,** and **Equation 11**. The radial velocity varies with respect to the guide vane position. The radial velocity is very low around the trailing edge of the guide vane, and it increases between the guide vanes. The variation for the full load condition ($\alpha_p = 1.3$) is high compared to the part load ($\alpha_p = 0.64$) and design load ($\alpha_p = 1$) conditions. Unlike radial velocity, the tangential velocity does not show a clear signature of periodic variation. The tangential velocity within the vaneless space is nearly similar, around 0.5, because it is dependent on the angular speed of the runner. The angular speed of the runner is similar for all load conditions. Contours of radial and tangential velocity across the guide vane and stay vane passages are extracted and shown in **Figure 31**. The colour scale of the contours is normalized with the maximum theoretical velocity, and the contours are shown for the part load condition in pump mode. The radial velocity around the guide vane is high as discussed before. Furthermore, high radial velocity vortices (marked as "Passage vortex") can be seen, which travel towards the stay vane passage through the guide vanes. The tangential velocity, as expected, is very high near to the runner and the vaneless space. We can also see high tangential velocity within the guide vane passage, and the vortex travels towards the stay vanes. Interestingly, the wake from the guide vane is clearly visible in the contours of the tangential velocity. An accelerating flow of high velocity between two guide vanes and the wake of low velocity interacts in the stay vane passage, which can be seen as mixed vortex shedding.

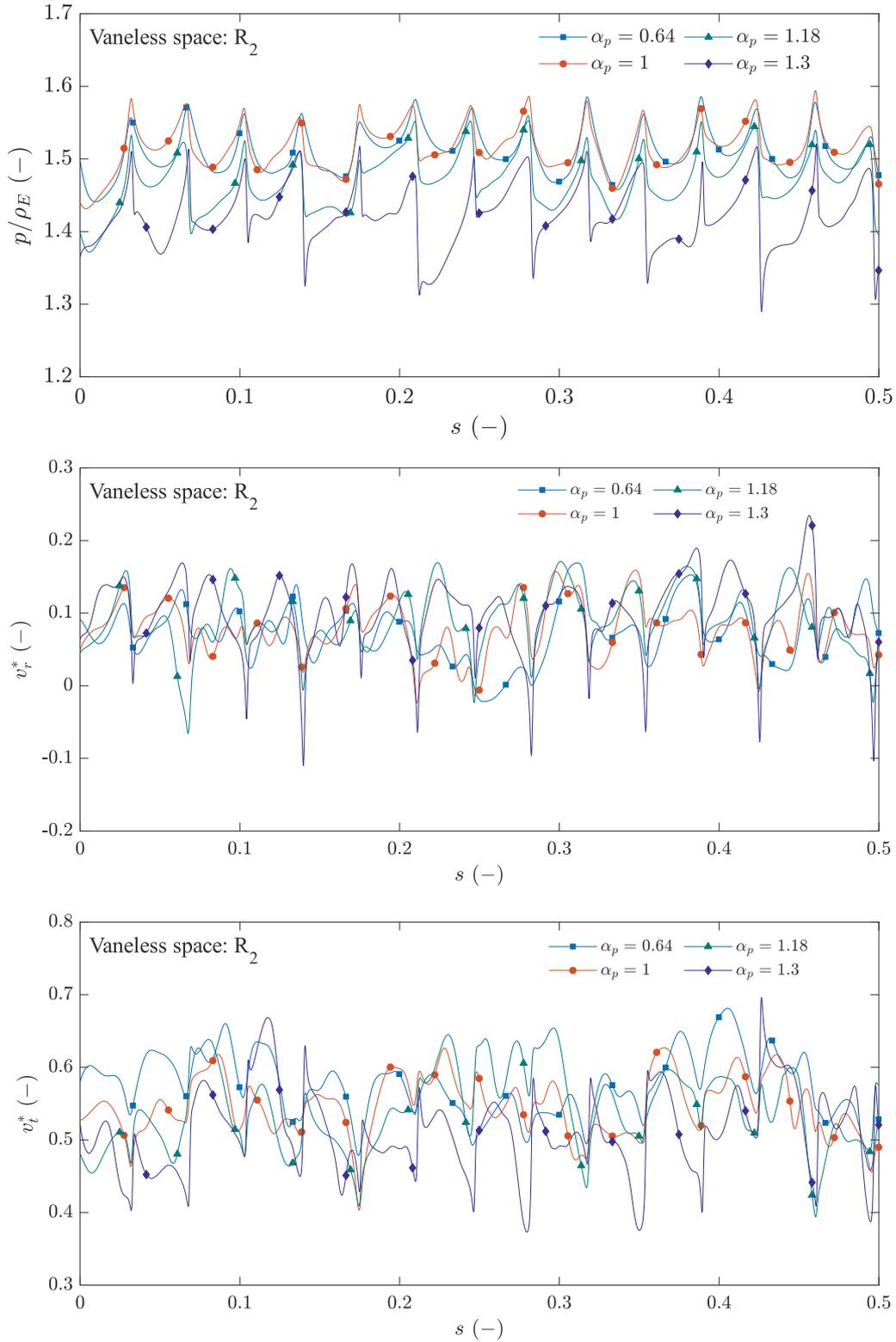

**Figure 30** Pressure and velocity variation in the vaneless space runner inlet for different load conditions. Pitch on x-axis corresponds to 0 – 180° circumference of the runner inlet.

C Trivedi

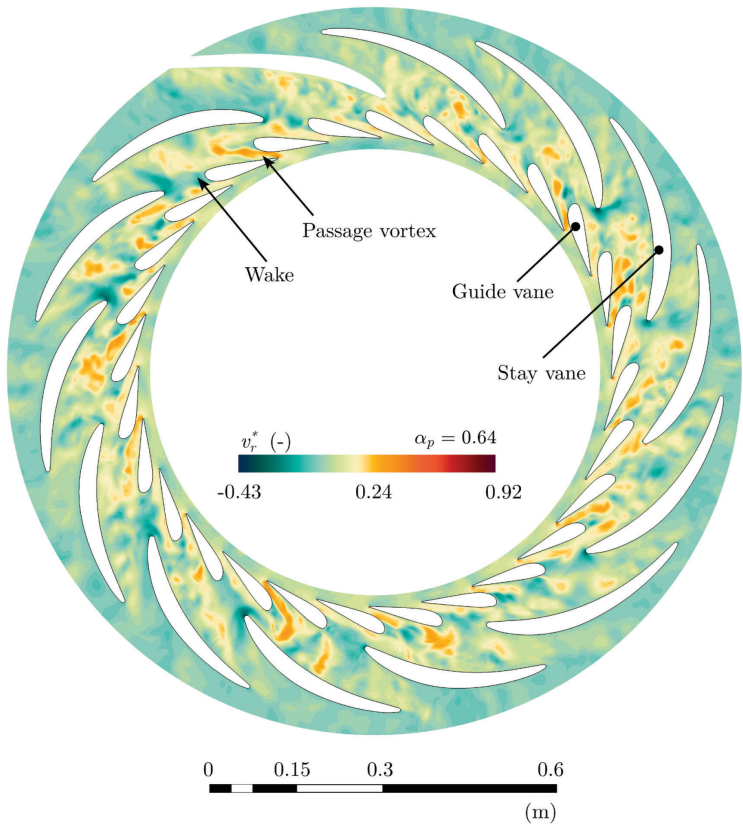

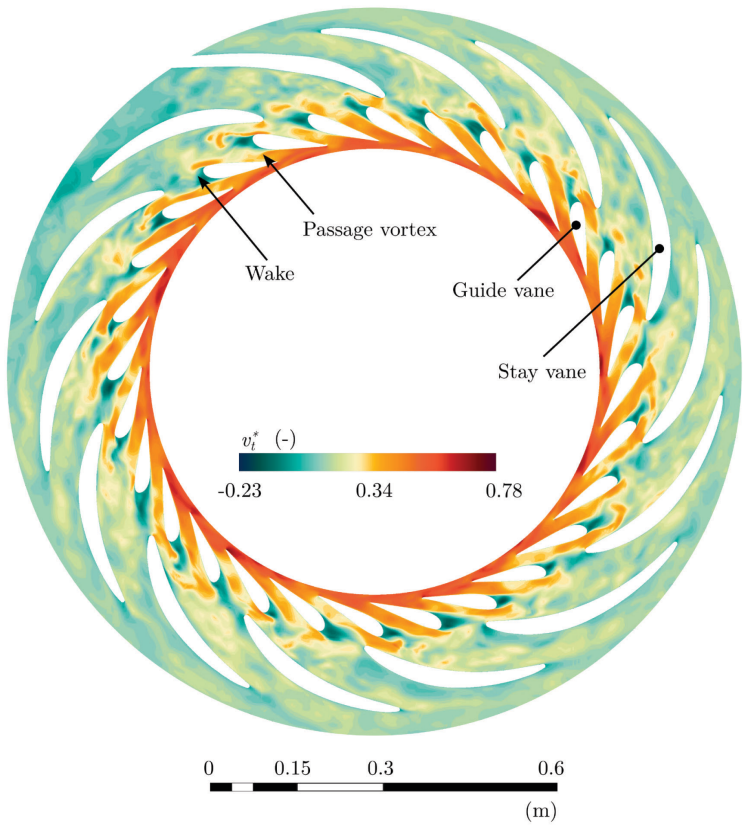

**Figure 31** Contour of radial and tangential velocity in the guide vane and stay vane passages at part load condition.

To investigate the time-dependent pressure fluctuations, the data from the three observation points ($\Phi_1$, $\Phi_2$, and $\Phi_3$) are analysed. The point $\Phi_1$ is placed between the stay vane and the guide vane passages; the point $\Phi_2$ is placed between the guide vanes; and the point $\Phi_3$ is placed in the vaneless space between the guide vanes and the runner inlet. Unsteady pressure fluctuations and the frequency spectra at these locations are shown in **Figure 32**. The fluctuations at $\Phi_1$ are both systematic and stochastic. The low frequency, small amplitude, fluctuations represent the vortex shedding and the wake effect from the guide vane leading edge. **Appendix B** shows the details about the frequencies and the amplitudes. High amplitude fluctuations represent the blade passing frequency. The frequency spectra show the blade passing frequency ($f_b = 6$) and the harmonics clearly. Similarly, at $\Phi_2$ location, the fluctuations correspond to the blade passing frequency. At this location, the amplitude of the blade passing frequency is maximum as compared to the other locations in the vaneless space. Interestingly, the harmonics of the blade passing frequency are up to nine, which are not observed at other locations in the turbine. In the vaneless space location, $\Phi_2$, the pressure fluctuations are largely related to the blade passing frequency, though the signature of the fluctuations is different from that of the locations $\Phi_1$ and $\Phi_2$. When comparing to the other load conditions of the turbine, no large variation in the signature of the pressure fluctuations is observed. Therefore, analysis of those data is not presented.

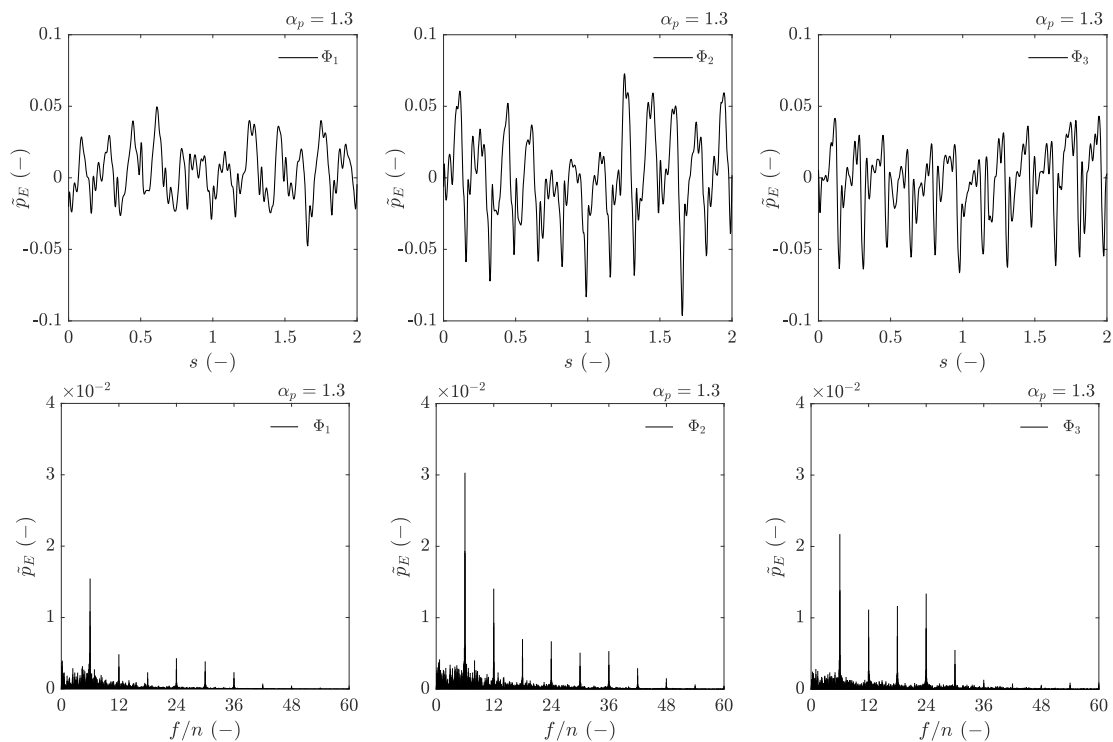

**Figure 32** Unsteady pressure fluctuations and frequency spectra in the guide vane passage at full load condition.

# Conclusions

We investigated the reversible pump-turbine at design and off design conditions for both turbine and pump modes operation. The computational domain consisted of all components of the turbine, including a spiral casing with extended inlet, stay vanes, guide vanes, runner and an elbow type draft tube. Both steady state and unsteady state simulations were conducted. The computational domain included 58.19 million hexahedral nodes. Estimated total numerical error was 7.7% at design load condition. For detailed analysis of the flow characteristics, 21 observation points were created in the computational domain, enabling time-dependent pressure fluctuations. Three operating conditions in turbine mode and four operating conditions in pump mode were investigated.

**Turbine mode**

- The static pressure, radial, and tangential velocity variation is strongly dependent on the exact position of the blades relative to the guide vanes in the vaneless space.
- The results clearly show that the high amplitude frequency can be up to six times the fundamental frequency of the blade.
- Pressure gradient on the blade leading edge is 0.02 - 1.12 times the specific hydraulic energy, indicating the significant change of pressure on small radii of leading edge may induce very high stresses.
- Investigation of flow field in the blade passage shows high separation of flow on the suction side of the blade.
- The spectral analysis of pressure data at the observation points in the runner clearly indicates up to 30 harmonics of the runner rotational frequency. This frequency also coincides with the guide vane passing frequency.
- Flow field analysis in the draft tube at part load shows the presence of several swirling zones having local rotation; flow in the centre of the draft tube is pumping towards the runner.

**Pump mode**

Flow direction in the pump mode operation reverses, where the flow enters the draft tube and leaves from the spiral casing.

- Investigation of the flow field in the draft tube revealed interesting characteristics. Secondary flow in the draft tube develops the Dean vortices up the runner edge.
- Three distinct zones of swirling flow were obtained in the draft tube. (1) Outer edge swirl — zone near to the wall, has high rotational velocity and travels radially inward. (2) core swirl — zone in the centre of the draft tube travels radially outward towards the wall and collides with the outer edge swirl. (3) Dean vortices — two vortices spin in opposite directions on the

- draft tube elbow axis. The Dean vortices draw momentum from the outer edge swirl and mix with the core swirl.
- The blade pressure loading increases from the trailing edge to the leading edge.
- Frequency spectra in the runner revealed high amplitude and wide band low frequencies indicating a complex and stochastic flow field. Furthermore, the spectra show frequencies of runner rotation, guide vane passing, and the blade passing. Presence of blade passing frequency in the runner is uncommon.
- Contour of velocity field revealed several vortical zones and inter-blade vortices in the blade channel.
- Flow field in the vaneless space, guide vane passage, and the stay vane passage revealed strong interactions among the blade leading edge vortex shedding and the guide vane trailing edge stagnation point. Furthermore, a vortex of high radial velocity travels through the guide vanes and interacts with the leading edge wake of the guide vanes.

The frequency spectra in the guide vanes showed amplitudes of frequencies up to tenth harmonics of the blade passing frequency, indicating the strong presence of rotor-stator interaction frequencies in the turbine. The frequencies can potentially align with the higher mode eigen frequencies and induce resonance. The combination of guide vane passing frequency, blade passing frequency (higher harmonics), runner rotational frequency, and the eigen frequency potentially results in strong flow-induced resonance in the turbine and eventually fatigue failure.

# Acknowledgements

The numerical simulations of reversible pump-turbine have been conducted using high performance computing cluster—IDUN at NTNU, Norway.

# Funding

This research has received financial support from the European Union's Horizon Europe "Research and Innovation Action" programme HORIZON-CL5-2023-D3-01-13 under the grant agreement number 101136176. Project: Novel long-term electricity storage technologies for flexible hydropower – store2hydro.

# Disclaimer

Funded by the European Union. Views and opinions expressed are however those of the author(s) only and do not necessarily reflect those of the European Union or European Commission. Neither the European Union nor the granting authority can be held responsible for them.

# Data availability statement

The data that support the findings of this study are openly available in DataverseNO at http://doi.org/10.18710/ECWM4E, reference number [37].

# Preprint

Author's Original Manuscript (AOM) is submitted to arXiv preprint server (https://arxiv.org/). The temporary submission number is: submit/7178242.

# Notation

$c$ = Chord of the blade (m)
$D$ = Runner outlet diameter (m)
$E$ = Specific hydraulic energy (J kg$^{-1}$), $E = gH$
$\hat{e}$ = Error (%)
$f$ = Frequency (Hz)
$H$ = Head (m)
$l$ = Length (m)
$n$ = Runner angular speed (o min$^{-1}$)
$n_{ED}$ = Speed factor (-), $nD/\sqrt{E}$
$p$ = Pressure (Pa)
$Q$ = Flow rate (m$^3$ s$^{-1}$)
$Q_{ED}$ = Discharge factor (-), $QD^2/\sqrt{E}$
$r$ = Radius (m); $r_{ref}$ = 0.1745 m
$s$ = Runner pitch (-)
$T$ = Torque (Nm)
$t$ = Time (s)
$u$ = Velocity (m s$^{-1}$)
$v$ = Velocity (m s$^{-1}$)

$v_c$ = Characteristic velocity (m s$^{-1}$), $\sqrt{2gH}$

$w$ = Velocity (m s$^{-1}$)

$x, y, z$ = Spatial coordinates

**Greek letter**

$\alpha$ = Guide vane angle / turbine load (-)

$\rho$ = Density (kg m$^{-3}$)

$\eta$ = Efficiency (-)

$\Phi$ = Observation points in guide vane passage

$\psi$ = Observation points in draft tube

$\Pi$ = Observation points in runner

$\Lambda$ = Observation points at turbine inlet conduit

**Subscript**

b = Blade

exp = Experimental

ext = Extrapolation

gv = Guide vane

h = Hydraulic

num = Numerical

p = Pump mode

r = Radial

t = Turbine mode, total, tangential

v = Validation

# APPENDIX A: Amplitude of rotor-stator interaction in turbine mode

**Table A1** Amplitude $\tilde{p}_E$ (%) of the blade passing frequency and harmonics at $\alpha_t = 0.64$.

| Location | $f_b$ | $2f_b$ | $3f_b$ | $4f_b$ | $5f_b$ |
|---|---|---|---|---|---|
| $\Lambda_1$ | 0.1082 | 0.0094 | 0.0101 | 0.0312 | 0.0512 |
| $\Lambda_2$ | 0.1081 | 0.0093 | 0.0100 | 0.0308 | 0.0523 |
| $\Phi_1$ | 0.7327 | 0.1157 | 0.0801 | 0.1333 | 0.1212 |
| $\Phi_2$ | 2.3641 | 0.6751 | 0.2963 | 0.2595 | 0.1966 |
| $\Phi_3$ | 2.4126 | 0.7491 | 0.3877 | 0.4386 | 0.1761 |
| $\Psi_1$ | 0.1461 | 0.0844 | 0.0277 | 0.0245 | 0.0084 |
| $\Psi_5$ | 0.1168 | 0.0239 | 0.0107 | 0.0128 | 0.0065 |

**Table A2** Amplitude $\tilde{p}_E$ (%) of the guide vane passing frequency and harmonics at $\alpha_t = 0.64$.

| Location | $0.5f_{gv}$ | $f_{gv}$ | $2f_{gv}$ |
|---|---|---|---|
| $\Pi_1$ | 1.1684 | 2.3273 | 0.1551 |
| $\Pi_3$ | 0.0289 | 1.0948 | 0.0948 |
| $\Pi_4$ | 0.0180 | 0.4415 | 0.0230 |
| $\Pi_5$ | - | 0.1991 | - |
| $\Pi_6$ | 0.0263 | 0.5795 | 0.0260 |
| $\Pi_7$ | 0.0155 | 0.3319 | 0.0133 |
| $\Pi_8$ | - | 0.0689 | - |
| $\Pi_2$ | - | 0.0238 | - |

**Table A3** Amplitude $\tilde{p}_E$ (%) of the blade passing frequency and harmonics at $\alpha_t = 1$.

| Location | $f_b$ | $2f_b$ | $3f_b$ | $4f_b$ | $5f_b$ |
|---|---|---|---|---|---|
| $\Lambda_1$ | 0.2346 | 0.0993 | 0.0337 | 0.0840 | 0.1689 |
| $\Lambda_2$ | 0.2338 | 0.0993 | 0.0327 | 0.0821 | 0.1696 |
| $\Phi_1$ | 1.8056 | 0.4617 | 0.2383 | 0.3017 | 0.2647 |
| $\Phi_2$ | 4.8703 | 1.7832 | 0.7982 | 0.5299 | 0.3496 |
| $\Phi_3$ | 5.4221 | 2.1873 | 1.0949 | 0.8258 | 0.3878 |
| $\Psi_1$ | 0.3984 | 0.1025 | 0.0898 | 0.0592 | 0.0107 |
| $\Psi_5$ | 0.4027 | 0.0773 | 0.0350 | 0.0144 | 0.0053 |

**Table A4** Amplitude $\tilde{p}_E$ (%) of the guide vane passing frequency and harmonics at $\alpha_t = 1$.

| Location | $0.5f_{gv}$ | $f_{gv}$ | $2f_{gv}$ |
|---|---|---|---|
| $\Pi_1$ | 2.1240 | 5.6226 | 0.1821 |
| $\Pi_3$ | 0.1434 | 1.9763 | 0.1117 |
| $\Pi_4$ | 0.0588 | 0.5438 | 0.0261 |
| $\Pi_5$ | - | 0.2598 | 0.0139 |
| $\Pi_6$ | 0.0426 | 0.7497 | 0.0383 |
| $\Pi_7$ | 0.0651 | 0.3366 | 0.0112 |
| $\Pi_8$ | 0.0502 | 0.0654 | - |
| $\Pi_2$ | - | 0.0256 | - |

**Table A5** Amplitude $\tilde{p}_E$ (%) of the blade passing frequency and harmonics at $\alpha_t = 1.18$.

| Location | $f_b$ | $2f_b$ | $3f_b$ | $4f_b$ | $5f_b$ |
|---|---|---|---|---|---|
| $\Lambda_1$ | 0.3061 | 0.1420 | 0.0484 | 0.1062 | 0.1163 |
| $\Lambda_2$ | 0.3062 | 0.1415 | 0.0483 | 0.1045 | 0.1187 |
| $\Phi_1$ | 2.2847 | 0.6156 | 0.3176 | 0.4388 | 0.2452 |
| $\Phi_2$ | 5.4678 | 2.0115 | 0.9798 | 0.7230 | 0.3058 |
| $\Phi_3$ | 6.2548 | 2.5749 | 1.3223 | 0.9043 | 0.7684 |
| $\Psi_1$ | 0.6618 | 0.1881 | 0.1601 | 0.0471 | 0.0110 |
| $\Psi_5$ | 0.7127 | 0.0814 | 0.0371 | 0.0234 | - |

**Table A6** Amplitude $\tilde{p}_E$ (%) of the blade passing frequency and harmonics at $\alpha_t = 1.18$.

| Location | $0.5f_{gv}$ | $f_{gv}$ | $2f_{gv}$ |
|---|---|---|---|
| $\Pi_1$ | 2.2338 | 6.3684 | 0.6753 |
| $\Pi_3$ | 0.1520 | 1.9043 | 0.1367 |
| $\Pi_4$ | 0.0362 | 0.7518 | 0.0649 |
| $\Pi_5$ | 0.0146 | 0.3492 | 0.0384 |
| $\Pi_6$ | 0.0334 | 0.9059 | 0.0408 |
| $\Pi_7$ | 0.0673 | 0.5269 | 0.0366 |
| $\Pi_8$ | 0.0815 | 0.0958 | 0.0197 |
| $\Pi_2$ | 0.0421 | 0.0350 | 0.0120 |

# APPENDIX B: Amplitude of rotor-stator interaction in pump mode

**Table B1** Amplitude $\tilde{p}_E$ (%) of the blade passing frequency and harmonics at $\alpha_p = 0.64$.

| Location | $f_b$ | $2f_b$ | $3f_b$ | $4f_b$ | $5f_b$ |
|---|---|---|---|---|---|
| $\Lambda_1$ | - | - | - | - | - |
| $\Lambda_2$ | - | - | - | - | - |
| $\Phi_1$ | 1.1133 | 0.2108 | - | 0.1981 | 0.2999 |
| $\Phi_2$ | 2.8302 | 0.4351 | 0.4697 | 0.4034 | 0.5252 |
| $\Phi_3$ | 1.5706 | 0.6502 | 0.7527 | 0.9193 | 0.3552 |
| $\Psi_1$ | - | - | - | - | - |
| $\Psi_5$ | - | - | - | - | - |

**Table B2** Amplitude $\tilde{p}_E$ (%) of the guide vane passing frequency and harmonics at $\alpha_p = 0.64$.

| Location | $0.5f_{gv}$ | $f_{gv}$ | $2f_{gv}$ |
|---|---|---|---|
| $\Pi_1$ | 0.3173 | 1.3294 | 0.2038 |
| $\Pi_3$ | 0.6842 | 1.6977 | 0.1305 |
| $\Pi_4$ | 0.2207 | 0.7488 | 0.0397 |
| $\Pi_5$ | 0.1697 | 0.3779 | 0.0139 |
| $\Pi_6$ | 0.2821 | 0.9928 | 0.0405 |
| $\Pi_7$ | 0.1513 | 0.5014 | 0.0202 |
| $\Pi_8$ | - | - | - |
| $\Pi_2$ | - | - | - |

**Table B3** Amplitude $\tilde{p}_E$ (%) of the blade passing frequency and harmonics at $\alpha_p = 1$.

| Location | $f_b$ | $2f_b$ | $3f_b$ | $4f_b$ | $5f_b$ |
|---|---|---|---|---|---|
| $\Lambda_1$ | - | - | - | - | - |
| $\Lambda_2$ | - | - | - | - | - |
| $\Phi_1$ | 0.9197 | 0.3550 | 0.1877 | 0.1360 | 0.4055 |
| $\Phi_2$ | 2.8605 | 1.1948 | 0.7943 | 0.3280 | 0.5960 |
| $\Phi_3$ | 2.0677 | 1.1122 | 1.0510 | 1.0863 | 0.3261 |
| $\Psi_1$ | - | - | - | - | - |
| $\Psi_5$ | - | - | - | - | - |

**Table B4** Amplitude $\tilde{p}_E$ (%) of the guide vane passing frequency and harmonics at $\alpha_p = 1$.

| Location | $0.5f_{gv}$ | $f_{gv}$ | $2f_{gv}$ |
|---|---|---|---|
| $\Pi_1$ | 0.8598 | 1.4002 | 0.1893 |
| $\Pi_3$ | 1.8181 | 1.4997 | 0.1906 |
| $\Pi_4$ | 0.4060 | 0.6888 | - |
| $\Pi_5$ | 0.2043 | 0.3384 | - |
| $\Pi_6$ | 0.6235 | 0.9637 | - |
| $\Pi_7$ | 0.3901 | 0.5107 | - |
| $\Pi_8$ | - | 0.1468 | - |
| $\Pi_2$ | 0.5425 | - | - |

**Table B5** Amplitude $\tilde{p}_E$ (%) of the blade passing frequency and harmonics at $\alpha_p = 1.3$.

| Location | $f_b$ | $2f_b$ | $3f_b$ | $4f_b$ | $5f_b$ |
|---|---|---|---|---|---|
| $\Lambda_1$ | - | - | - | - | - |
| $\Lambda_2$ | - | - | - | - | - |
| $\Phi_1$ | 1.5463 | 0.4858 | 0.2329 | 0.4304 | 0.3870 |
| $\Phi_2$ | 3.0305 | 1.4070 | 0.7006 | 0.6702 | 0.5107 |
| $\Phi_3$ | 2.1732 | 1.1152 | 1.1670 | 1.3398 | 0.5505 |
| $\Psi_1$ | - | - | - | - | - |
| $\Psi_5$ | - | - | - | - | - |

**Table B6** Amplitude $\tilde{p}_E$ (%) of the guide vane passing frequency and harmonics at $\alpha_p = 1.3$.

| Location | $0.5f_{gv}$ | $f_{gv}$ | $2f_{gv}$ |
|---|---|---|---|
| $\Pi_1$ | 0.9948 | 1.4813 | 0.2868 |
| $\Pi_3$ | 2.2099 | 0.9570 | 0.2474 |
| $\Pi_4$ | 0.6985 | 0.6332 | - |
| $\Pi_5$ | 0.3275 | 0.3646 | - |
| $\Pi_6$ | 0.6929 | 0.8791 | - |
| $\Pi_7$ | 0.3194 | 0.5234 | - |
| $\Pi_8$ | - | - | - |
| $\Pi_2$ | - | - | - |